\documentclass[11pt,twoside]{article}
\usepackage{latexsym}
\usepackage{amssymb,amsbsy,amsmath,amsfonts,amssymb,amscd}
\usepackage{epsfig, graphicx}
\usepackage{hyperref}
\newcommand{\commentout}[1]{}
\newcommand{\R}{\mathbb{R}}

\newcommand{\I}{{\rm 1\:\!\!\! I}}

\newcommand {\Chi} {{\bf \raise 2pt \hbox{$\chi$}} }
\newcommand {\f}   {\frac}
\newcommand {\p}   {\partial}

\newcommand {\proof} {\noindent {\bf Proof}. }
\newcommand{\beq}{\begin{equation}}
\newcommand{\eeq}{\end{equation}}
\newcommand{\bea} {\begin{array}{rl}}
\newcommand{\eea} {\end{array}}
\newcommand{\bepa}{\left\{ \begin{array}{l}}
\newcommand{\eepa} {\end{array}\right.}
\newtheorem{theorem}{Theorem}[section]
\newtheorem{lemma}[theorem]{Lemma}

\newtheorem{corollary}[theorem]{Corollary}
\newcommand{\qed}{{ \hfill
                    {\unskip\kern 6pt\penalty 500 \raise -2pt\hbox{\vrule\vbox to 6pt{\hrule width 6pt
                    \vfill\hrule}\vrule} \par}   }}
\title{\Large \bf Traveling plateaus for a hyperbolic Keller-Segel system with attraction and repulsion: existence and branching instabilities}
\author{Beno\^ \i t Perthame \thanks{UPMC, CNRS UMR 7598, Laboratoire Jacques-Louis Lions, F-75005, Paris and
INRIA Paris-Rocquencourt, Equipe BANG. Email: benoit.perthame@upmc.fr}
\and Christian Schmeiser \thanks{Institute for Mathematics, University of Vienna, Nordbergstra\ss e 15, 1090 Vienna, Austria and RICAM, Linz, Austria.
Email: christian.schmeiser@univie.ac.at}
\and Min Tang  \footnotemark[1] \thanks{
Email: tangmin1002@gmail.com}
\and Nicolas Vauchelet  \footnotemark[1] \thanks{Email: nicolas.vauchelet@upmc.fr}
}

\date{\today}

\begin{document}
\maketitle
\pagestyle{plain}
\pagenumbering{arabic}

\begin{abstract}
How can repulsive and attractive forces, acting on a conservative system, create stable traveling patterns or branching instabilities?
We have proposed to study this question in the framework of the hyperbolic Keller-Segel system with logistic sensitivity. This is a model system motivated by experiments on cell communities auto-organization, a field which is also called socio-biology. We continue earlier modeling work, where we have shown numerically that branching patterns arise for this system and we have analyzed this instability by formal asymptotics for small diffusivity of the chemo-repellent.

Here we are interested in the more general situation, where the diffusivities of both the chemo-attractant and
the chemo-repellent are positive. To do so, we develop an appropriate functional analysis framework. We apply our method to two cases. Firstly we analyze steady states. Secondly we analyze traveling waves when neglecting the degradation coefficient of the chemo-repellent; the unique wave speed appears through a singularity cancelation which is the main theoretical difficulty. This shows that in different situations the cell density takes the shape of a plateau.

The existence of steady states and traveling plateaus are a symptom of how rich the system is and why branching instabilities can occur. Numerical tests show that large plateaus may split into smaller ones, which remain stable.

\end{abstract}

\\[2mm]
\noindent {\bf AMS Subjects Class.} 35L45, 35L67, 65M99, 92C17

\section{Introduction}
\label{sec:int}

We study a form of the hyperbolic Keller-Segel system with logistic sensitivity for a chemoattractant and
constant sensitivity for a chemo-repellent, given by the set of equations
\beq\label{eq:HYKS}
\left\{\begin{array}{l}
\p_tn+\mbox{div}\left[\mu_c n\left(1-\frac{n}{n_{max}}\right)\nabla c - \mu_S n\nabla S\right] = 0 \,,\\[2mm]
-D_c\Delta c+\frac{c}{\tau_c} = \alpha_c n \,,\\[2mm]
\p_tS - D_S\Delta S + \frac{S}{\tau_S} = \alpha_S n \,.
\end{array}\right.
\eeq
This system is reminiscent of a large class of models used with biomedical motivations to represent the auto-organization of cells that are able to produce attractive and repulsive chemicals \cite{GKCB,PMO}, but also in other areas of population dynamics \cite{BMP,MRS}. The logistic sensitivity $\mu_c(1-n/n_{max})$ takes into account a volume filling (or quorum sensing) effect, i.e., a reduction of the cell response to the chemo-attractant (whose concentration is denoted by $c$), which prevents overcrowding \cite{HPsurvey,PH, perth1}. The special form at hand has been proposed in \cite{CPSTV} as a reduced model for a more detailed system to study complex patterns as the
dendritic ramification of \emph{Bacillus subtilis}, recently obtained with high nutrient experiments in \cite{JOHS,JOHS1,MHHPSP}, whereas pattern formation based on local nutrient depletion is also possible \cite{MSM,GKCB,murray}. It includes a chemo-repellent of concentration $S$ (that can be interpreted as the effect of surfactin) with a constant sensitivity $\mu_S$. The model neglects diffusion (i.e. random motion) of the cells. This has been proved to be mathematically correct, due to the logistic sensitivity, in \cite{DS,DaPe} (see also \cite{BDD} for an earlier work).
Reaction-diffusion models are used for the chemicals with a quasi-stationarity assumption for the chemo-attractant. The
chemicals diffuse with diffusivities $D_c$ and $D_S$, they are degraded with relaxation times $\tau_c$ and $\tau_S$,
and they are produced by the cells with rates $\alpha_c$ and $\alpha_S$.

From \cite{BDS,BDD,DS} we know that, when $S\equiv 0$, the nonlinear term $n(1-n/n_{max})$
causes sharp fronts that connect alternatively the states $n=0$ and $n=n_{max}$ (see \cite{Hillen} for this terminology). The repellent force $\nabla S$ can generate surprising dynamics of the plateaus and branching instabilities may occur. This was shown in \cite{CPSTV}, and the instability could be analyzed for $D_S$ small, because the limiting system
with $D_S=0$ can be recast as an hyperbolic system according to a method introduced in \cite{LW}. Then, stability/instability of discontinuities can be seen as a transition from shock to rarefaction waves.

In the present paper, we are interested in the existence and branching instabilities of traveling plateau
solutions of (\ref{eq:HYKS}) for general diffusion coefficients. These are naturally obtained in one dimension and therefore we focus on the nondimensionalized system
\begin{equation}\label{eq:HYKS1d}\left\{
\begin{array}{ll}
\partial_tn+\partial_x \big(n(1-n)\partial_x c-n\partial_x S \big)=0 \,,\\[2mm]
-D_c\partial_x^2 c+  c = \alpha_c n \,,\\[2mm]
\partial_tS -D_s\partial_x^2 S+ S= \alpha_S n \,.
\end{array}
\right.
\end{equation}
For the nondimensionalization, $\tau_S$ has been chosen as the reference time and $n_{max}$ as the reference cell
density. The reference values for the chemical densities have been chosen such that the scaled versions of the
sensitivities $\mu_c$ and $\mu_S$ are equal to 1. All quantities in (\ref{eq:HYKS1d}) are dimensionless.
This is also true for the parameters $D_c$, $D_S$, $\alpha_c$, and $\alpha_S$, where we point out that
the scaled diffusivities $D_c$ and $D_S$ actually represent (in unscaled notation) $D_c \tau_c/l^2$ and
$D_S \tau_S/l^2$, respectively, where $l$ is the reference length.

We develop a functional analytic framework for the study of traveling waves. Since it is simpler to introduce it for steady states of (\ref{eq:HYKS1d}), we address this issue in Section \ref{sec:steady}. We prove the existence of a family of steady states characterized by the size of the plateau, when it is small enough, or when $D_S$ is close enough to $D_c$ (Section \ref{sec:stst2}). Numerical tests illustrate that the smallness condition is necessary for stability. In Section \ref{sec:tp}, we
show that steady states are replaced by traveling plateaus, when degradation of the chemo-repellent is neglected
(i.e. $\tau_S = \infty$ in (\ref{eq:HYKS})). The method is an extension of the functional analytic framework for steady states, where the propagation speed is determined naturally by a singularity analysis. The total number of cells (or the size of the plateau) defines a family of traveling waves with different speeds.

\section{Steady states with small total mass}
\label{sec:steady}

The existence of stationary states for system \eqref{eq:HYKS1d} is both the most natural question and the easiest to illustrate the method we use throughout the paper. Therefore we begin with this issue. We first state a theoretical result that involves a smallness condition. Then, we confirm with numerical results that this smallness condition is necessary.

We shall look for solutions of the steady state system
\begin{equation}\label{laststst}\left\{
\begin{array}{ll}
\partial_x \big(n(1-n)\partial_x c-n\partial_x S \big)=0 \,,\\[2mm]
-D_c\partial_x^2 c+  c = \alpha_c n \,,\\[2mm]
-D_s\partial_x^2 S+ S= \alpha_S n \,,
\end{array}
\right.
\end{equation}
complemented with the boundary conditions
\beq\label{eq:boundsteady}
n(\pm\infty)=c(\pm\infty)=s(\pm\infty)=0 \,.
\eeq
Integration of the first equation then gives vanishing flux:
$$
n \big[(1-n)\partial_x c-\partial_x S \big]=0 \,.
$$
We shall be interested in plateaus of the cell density with sharp boundaries, such that $n$ jumps between $n=0$ and
\beq\label{eq:ststn}
n=1 - \f{\partial_x S}{\partial_x c} > 0
\eeq
(satisfying the Rankine-Hugoniot jump conditions). The convexity of the flux function is
determined by the sign of $\p_x c$. The above jumps satisfy the entropy condition if either $n=0$ on the left, $n>0$
on the right of the jump, and $\p_x c > 0$ at the jump point; or $n>0$ on the left, $n=0$ on the right, and $\p_x c < 0$.

\begin{theorem}\label{th:laststst}
Assume that $\gamma := \f{\alpha_S D_c}{\alpha_c D_S}<1$.  Then, for $L$ small enough, there exists a unique
entropy solution of \eqref{laststst} of the form
$$
n(x)= \left\{ \begin{array}{ll}
0 & \text{for } \; x \notin [0,L],
\\[3mm]
1- \gamma + O(L) \quad & \text{for } \; x \in (0,L),
\end{array}\right.
$$
with $n\in C(0,L)$ and symmetric around $x=L/2$.
\end{theorem}
For the existence of steady state plateau solutions, we need the effect of the chemo-attractant to dominate
the effect of the chemo-repellent. In terms of the original unscaled parameters,
$\gamma = \frac{\alpha_S\mu_S}{D_S}\,\frac{D_c}{\alpha_c\mu_c}$ holds. Thus, the strength of the attractive (respectively repulsive) effect is measured by the product of the production rate and the sensitivity divided by the
diffusivity of the chemical.

One can understand the occurrence of the free parameter $L$ as the result of mass conservation in the
dynamics \eqref{eq:HYKS1d}. Supposedly there is a one-to-one relation between $L$ and the total number of cells.
This is also the way plateaus of different size are produced in the numerical examples below.

\subsection*{Proof of Theorem \ref{th:laststst}}

The difficulty in using (\ref{eq:ststn}) for the computation of the cell density is to control the points where
$\partial_x c$ vanishes. By the symmetry assumption, one such point is at $x=L/2$, and we shall prove
that for $n$ of the form given in the theorem, it is the only one.

It will be convenient to rescale the problem by $x \to Lx$. Then, the support of $n$ is given by $[0,1]$,
where (\ref{eq:ststn}) still holds, and
\beq\label{eq:cs}
\left\{
\begin{array}{ll}
-\f{D_c}{L^2} \partial_x^2 c+  c = \alpha_c n \,,\\[2mm]
-\f{D_S}{L^2} \partial_x^2 S+ S= \alpha_S n \,.
\end{array}
\right.
\eeq
With the boundary conditions (\ref{eq:boundsteady}), explicit representations of $\p_xc$ and $\p_xS$ can be
computed for $x\in \R$:
\begin{eqnarray}
\partial_xc&=&-\f{\alpha_cL_c^2}{2}\left(\int_0^xe^{L_c(y-x)}n(y)dy
-\int_x^1e^{L_c(x-y)}n(y)dy\right) \,,\label{eq:soluLc1}
\\
\partial_xS&=&-\f{\alpha_SL_S^2}{2}\left(\int_0^xe^{L_S(y-x)}n(y)dy
-\int_x^1e^{L_S(x-y)}n(y)dy\right) \,,\label{eq:soluLs1}
\end{eqnarray}
with $L_S= L/\sqrt{D_S}$, $L_c=L/\sqrt{D_c}$. Thus the formula \eqref{eq:ststn} can be seen as the fixed point
equation
\beq \label{eq:fp}
n= {\cal F}[n] := 1- \gamma  \f{F[L_S,n]}{F[L_c,n]} \,,
\eeq
with
\beq\label{eq:F}
F[L_i,n](x) =-\int_{0}^x e^{L_i(y-x)}n(y)dy+\int_x^1e^{L_i(x-y)}n(y)dy \,,\qquad 0 < x < 1 \,,\quad i = S,c \,.
\eeq
Using the symmetry at $x=1/2$, i.e., $n(x)=n(1-x)$ for $x\in(0,1)$,
we restrict to $x\in (0,1/2)$ and write
\beq\label{eq:decomp}
F[L_i,n](x) =  e^{L_i x}\int_x^{1/2} \left(e^{-L_i y}+e^{-L_i(1-y)}\right) n(y)dy
+ \left( e^{-L_i(1-x)} -e^{-L_i x} \right) \int_0^x e^{L_iy} n(y)dy \,.
\eeq
As the next step, the limit as $L_i\to 0$ is taken out, and the zero at $x=1/2$ is eliminated by
$$
\frac{F[L_i,n](x)}{1/2-x} = 2\bar n(x) + L_i f[L_i,n](x) \,,
$$
with
\begin{eqnarray*}
\bar n(x) &=& \frac{1}{1/2-x} \int_x^{1/2} n(y) dy \,,\\
f[L_i,n](x) &=& \frac{1}{1/2-x} \int_x^{1/2} \frac{e^{L_i(x-y)} + e^{L_i(x+y-1)} - 2}{L_i} n(y)dy
+ \frac{e^{-L_i(1-x)} -e^{-L_i x}}{L_i(1/2-x)} \int_0^x e^{L_iy} n(y)dy \,.
\end{eqnarray*}
Since $L=0$ implies $L_i=0$, the unique solution of (\ref{eq:fp}) in this case is given by
$$
n(x)=n_0 := 1-\gamma \,.
$$
We shall apply the Banach fixed point theorem in the ball
\beq
\label{eq:B}
B := \left\{ n\in C([0,1/2]) :\ \|n-n_0\|_\infty<\frac{1-\gamma}{2} \right\} \,,
\eeq
of the Banach space $C([0,1/2])$ (equipped with the supremum norm $\|\cdot\|_\infty$). The essential
observation is collected in the following result.

\begin{lemma}\label{lm:Lipschitz}
For $0<L_i \le \bar L$, the linear mapping $n\mapsto f[L_i,n]$ on $C([0,1/2])$ is bounded with a bound only
depending on $\bar L$.
\end{lemma}

\proof
It is easily seen that both
$$
\frac{e^{L_i(x-y)} + e^{L_i(x+y-1)} - 2}{L_i} \qquad\mbox{and}\qquad \frac{e^{-L_i(1-x)} -e^{-L_i x}}{L_i(1/2-x)}
$$
are uniformly bounded for $0\le x,y\le 1/2$, $0<L_i\le \bar L$ with a bound $C(\bar L)$, implying
$$
\| f[L_i,n] \|_\infty \le C(\bar L) \left( 1 + \frac{e^{\bar L/2}}{2} \right) \|n \|_\infty \,.
$$
\qed

\begin{corollary}
For $n\in B$ and for $L$ small enough (recalling $L_c = L/\sqrt{D_c}$, $L_S = L/\sqrt{D_S}$),
$$
 2 \bar n + L_c f[L_c,n] \ge \frac{1-\gamma}{2}
$$
and
$$
 n \mapsto {\cal F}[n] = 1 - \gamma + \gamma \frac{L_c f[L_c,n] - L_S f[L_S,n]}{2 \bar n + L_c f[L_c,n]}
$$
maps $B$ into itself and is a contraction with respect to $\|\cdot\|_\infty$ with a Lipschitz constant proportional to $L$.
\end{corollary}
This concludes the proof of Theorem \ref{th:laststst}.

\subsection*{Formal asymptotic expansion -- shape of the plateau}

In this section the first few terms in an asymptotic expansion of the solution of (\ref{eq:fp}) will be computed. This
will shed light on the shape of the non-constant correction of the cell density plateau. Some of the necessary
computations are rather lengthy and will only be outlined. We start with the Taylor expansion
$F[L_i,n] = F_0[n] + L_i F_1[n] + L_i^2 F_2[n] + O(L_i^3)$ with
$$
F_j[n](x) = -\int_0^x \frac{(y-x)^j}{j!} n(y)dy + \int_x^1 \frac{(x-y)^j}{j!} n(y)dy \,.
$$
This in turn leads to
$$
{\cal F}[n] = {\cal F}_0[n] + L {\cal F}_1[n] + L^2 {\cal F}_2[n] + O(L^3) \,,
$$
with
\begin{eqnarray*}
{\cal F}_0[n] &=& 1-\gamma \,,\qquad
{\cal F}_1[n] = \gamma\left( \frac{1}{\sqrt{D_c}} - \frac{1}{\sqrt{D_S}}\right) \frac{F_1[n]}{F_0[n]} \,,\\
{\cal F}_2[n] &=& \gamma\left( \frac{1}{D_c} - \frac{1}{D_S} \right)\frac{F_2[n]}{F_0[n]} +
 \frac{\gamma}{\sqrt{D_c}} \left( \frac{1}{\sqrt{D_S}} - \frac{1}{\sqrt{D_c}}\right) \frac{F_1[n]^2}{F_0[n]^2} \,.
\end{eqnarray*}
Substitution of the ansatz $n = n_0 + L n_1 + L^2 n_2 + O(L^3)$ into (\ref{eq:fp}), re-expansion, and equating
coefficients of powers of $L$ then leads to
\begin{eqnarray*}
n_0 &=& 1-\gamma \,,\\
n_1 &=& {\cal F}_1[n_0] = \frac{\gamma}{2} \left( \frac{1}{\sqrt{D_S}} - \frac{1}{\sqrt{D_c}}\right) \,,\\
n_2 &=& D{\cal F}_1[n_0]n_1 + {\cal F}_2[n_0] = \bar n_2 + \frac{\gamma}{6}\left( \frac{1}{D_c} - \frac{1}{D_S} \right)
     (x-1/2)^2 \,,
\end{eqnarray*}
where $D{\cal F}_1$ denotes the Frechet derivative of ${\cal F}_1$ and $\bar n_2$ is a (explicitly computable) constant.
The $O(L)$-correction term $n_1$ is constant. It is negative
for $D_c < D_S$ and positive for $D_c>D_S$. The first non-constant correction occurs at $O(L^2)$. It is convex for
$D_c<D_S$ and concave for $D_c>D_S$. This agrees qualitatively with the numerical results of Section \ref{sec:stst2}.

Finally, we mention that it is a standard procedure to extend our rigorous results in order to justify the asymptotic
expansion for $n$ in the sense that the error $O(L^3)$ can be estimated in $C([0,1/2])$ by $CL^3$.

\subsection*{Numerical experiments}
\label{sec:numtest}

We carried out numerical tests that illustrate the analytical results and indicate that large plateaus may be unstable,
depending on the relation between $D_c$ and $D_S$.

We obtained numerical steady states as the limit for large times of a modified dynamics (where also the chemo-repellent
is determined from a quasistationary problem), which we have chosen for its simplicity:
\begin{equation} \label{eq:red-dyn}
\left\{\begin{array}{ll}
\p_tn+\partial_x \big(n(1-n)\partial_x c-n\partial_x S \big)=0 \,,\\[2mm]
-D_c\partial_x^2 c+  c = \alpha_c n \,,\\[2mm]
-D_S\partial_x^2 S+ S= \alpha_S n \,.
\end{array}
\right.
\end{equation}
We discretize the hyperbolic equation for $n$ by the Enquist-Osher finite volume method, which is conservative and can capture the shocks on both sides of the plateaus (see \cite{bouchut} for a recent introduction to the subject). The elliptic equations for $c$ and $S$ are solved by a finite difference method.
We indeed obtained that after some transient the numerical solutions converge to a steady state.

\begin{figure}
\begin{center}
\includegraphics[width=0.3\textwidth]{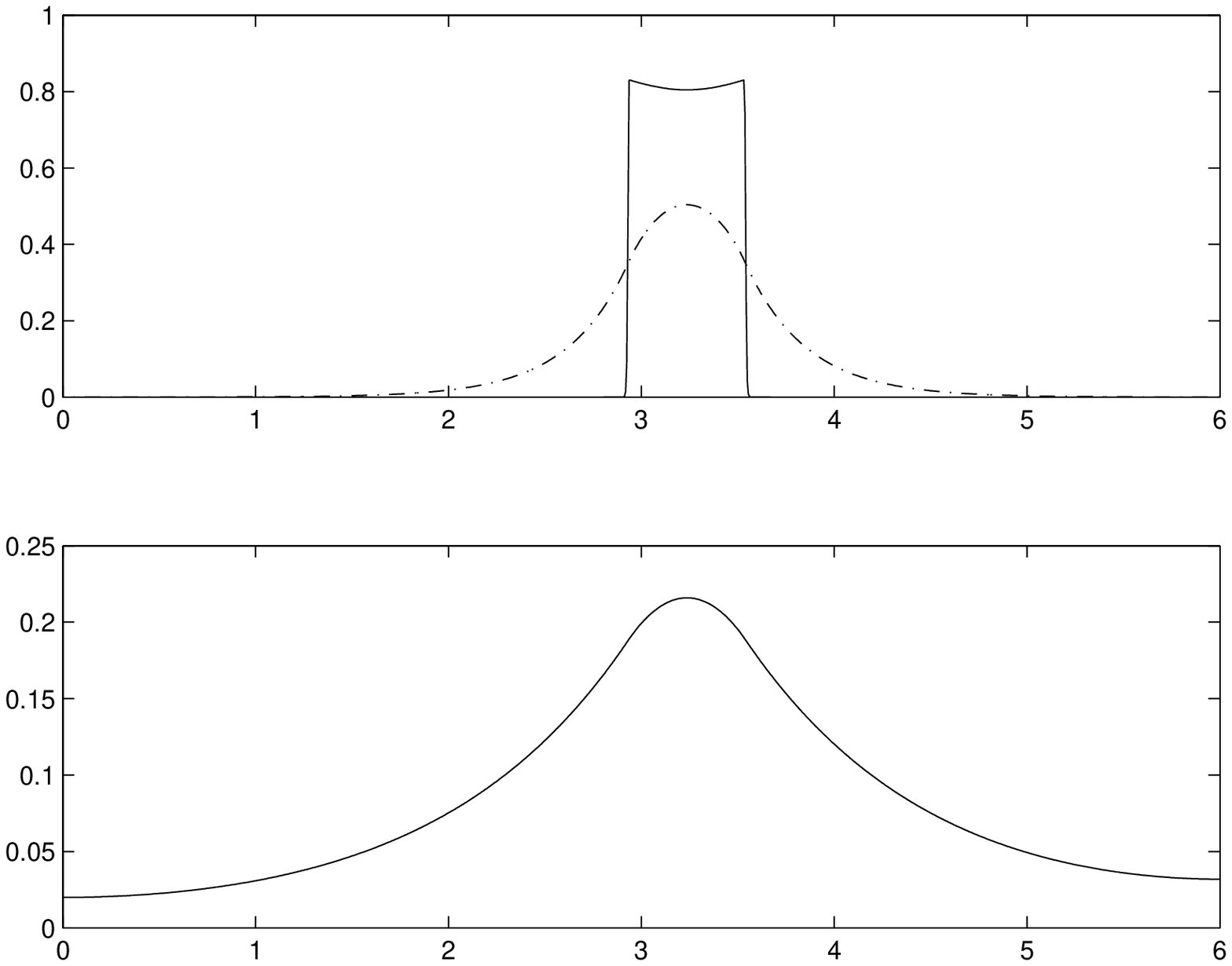}
\includegraphics[width=0.3\textwidth]{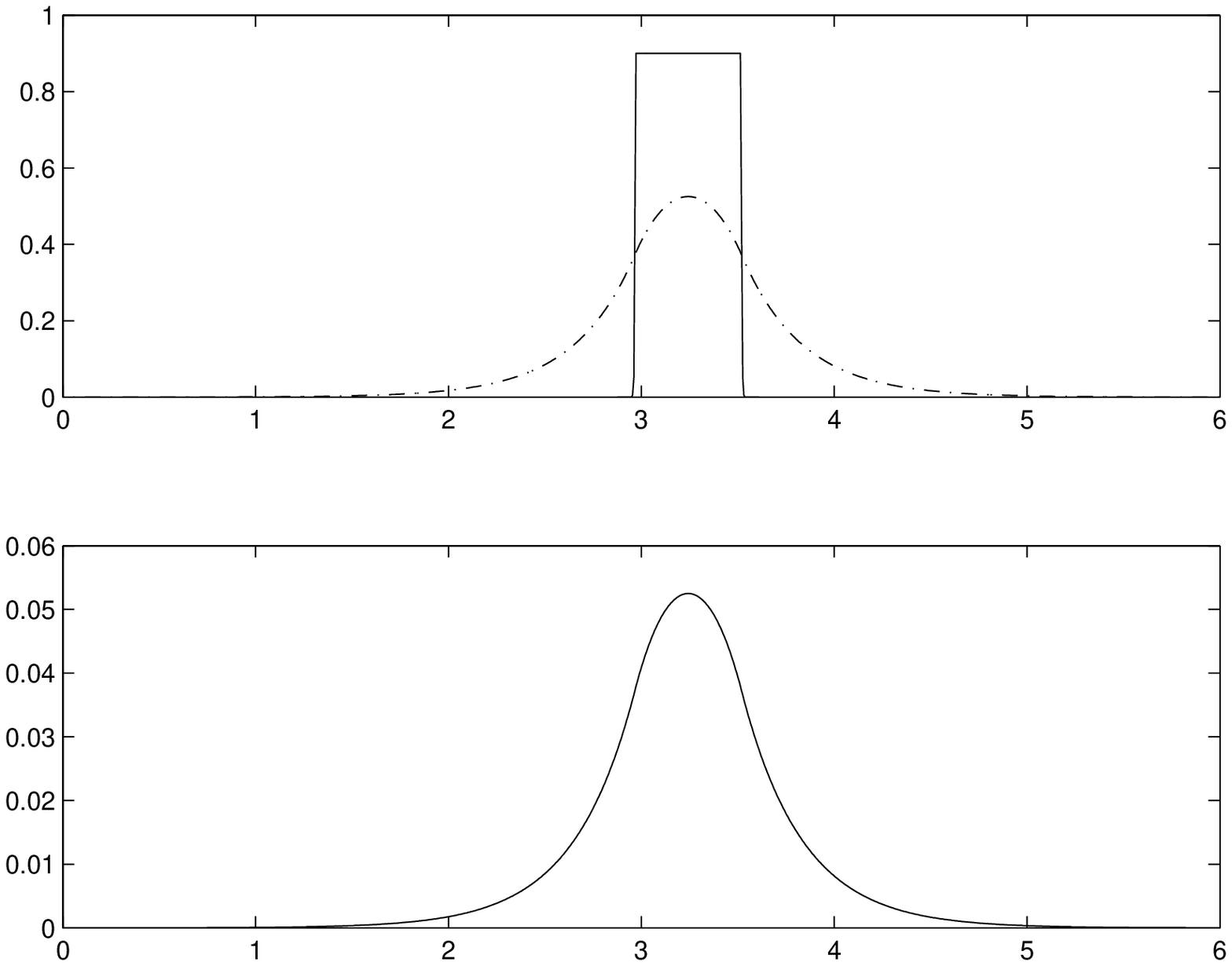}
\includegraphics[width=0.3\textwidth]{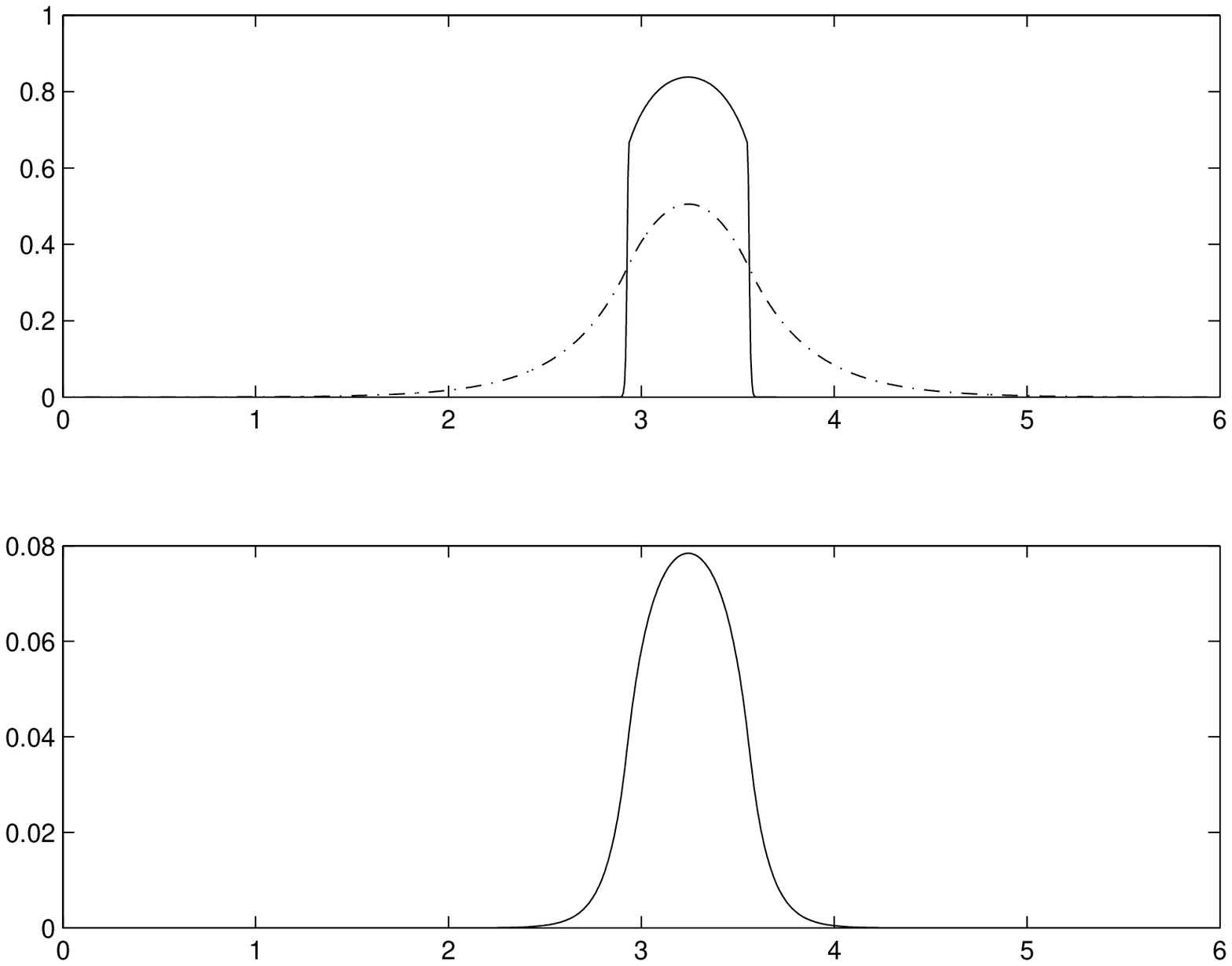}
\vspace{-2mm}
\caption{The steady state solution of \eqref{laststst} with total mass $M=0.5$ and different chemical production rates
and diffusivities. Left: $\alpha_c=1,\alpha_S=1,D_c=0.1,D_S=1$; middle: $\alpha_c=1,\alpha_S=0.1,D_c=0.1,D_S=0.1$; right: $\alpha_c=1,\alpha_S=0.1,D_c=0.1,D_S=0.01$.}
\label{fig:steady}
\end{center}
\end{figure}

\begin{figure}
\begin{center}
\includegraphics[width=0.45\textwidth]{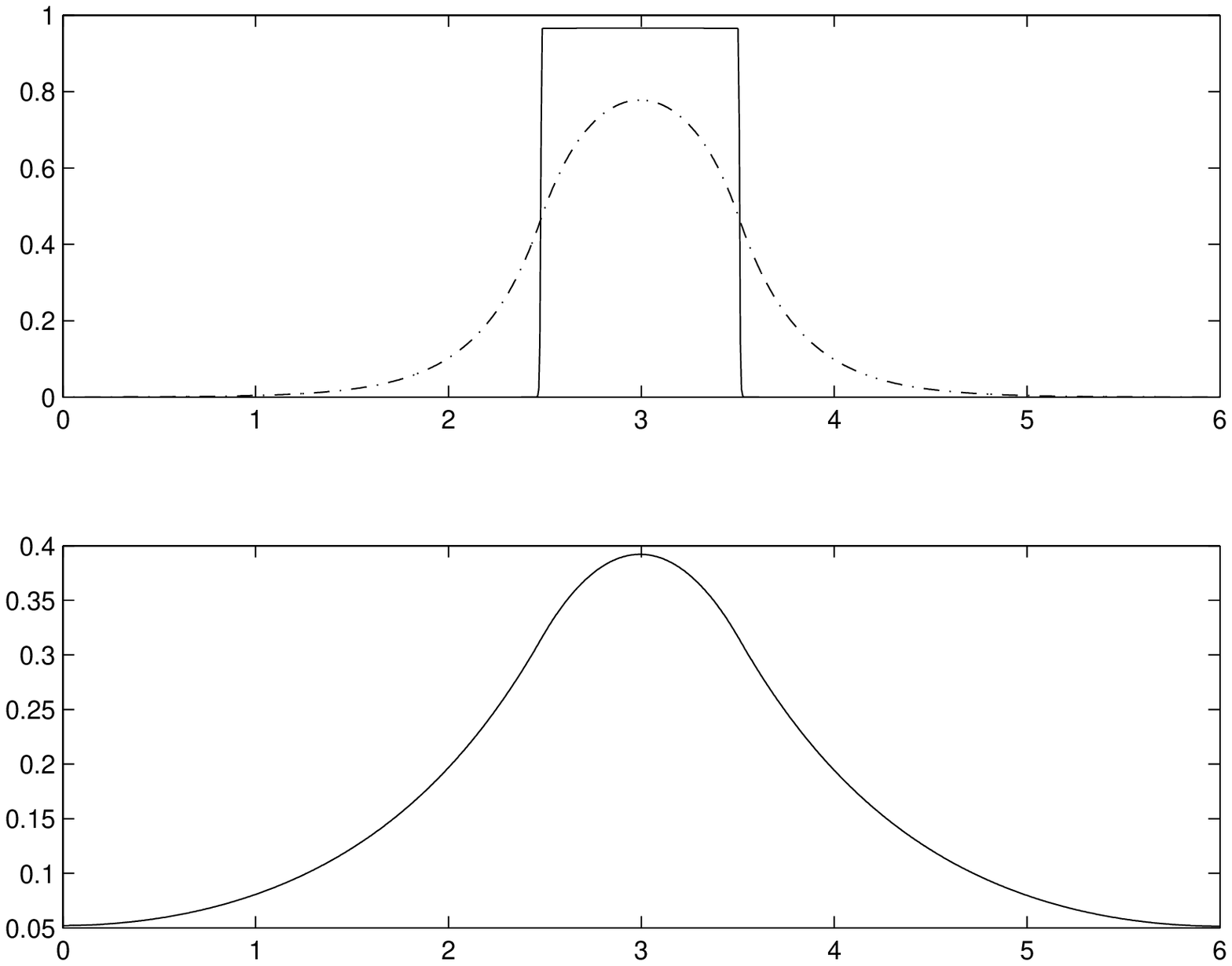}
\includegraphics[width=0.45\textwidth]{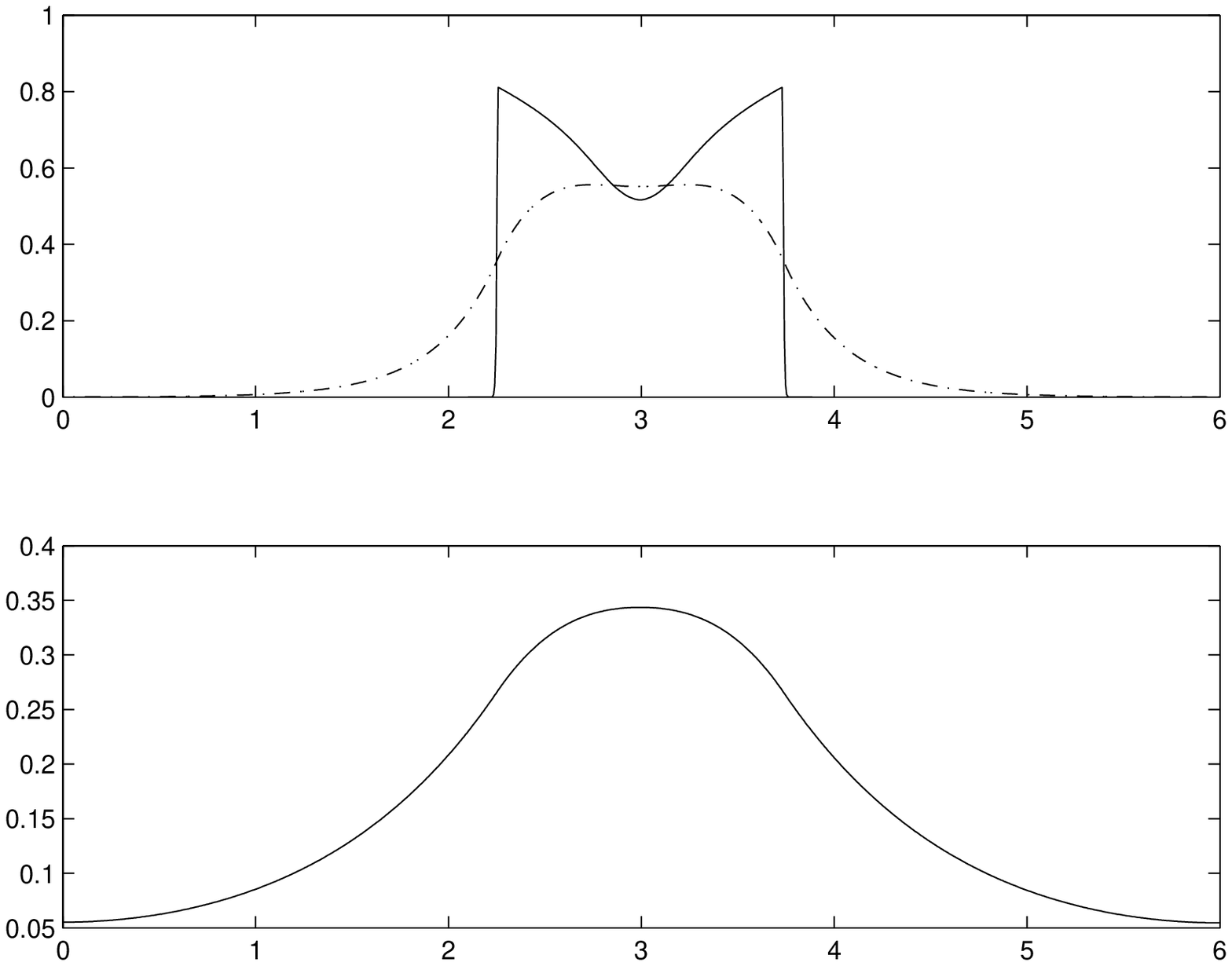}
\includegraphics[width=0.45\textwidth]{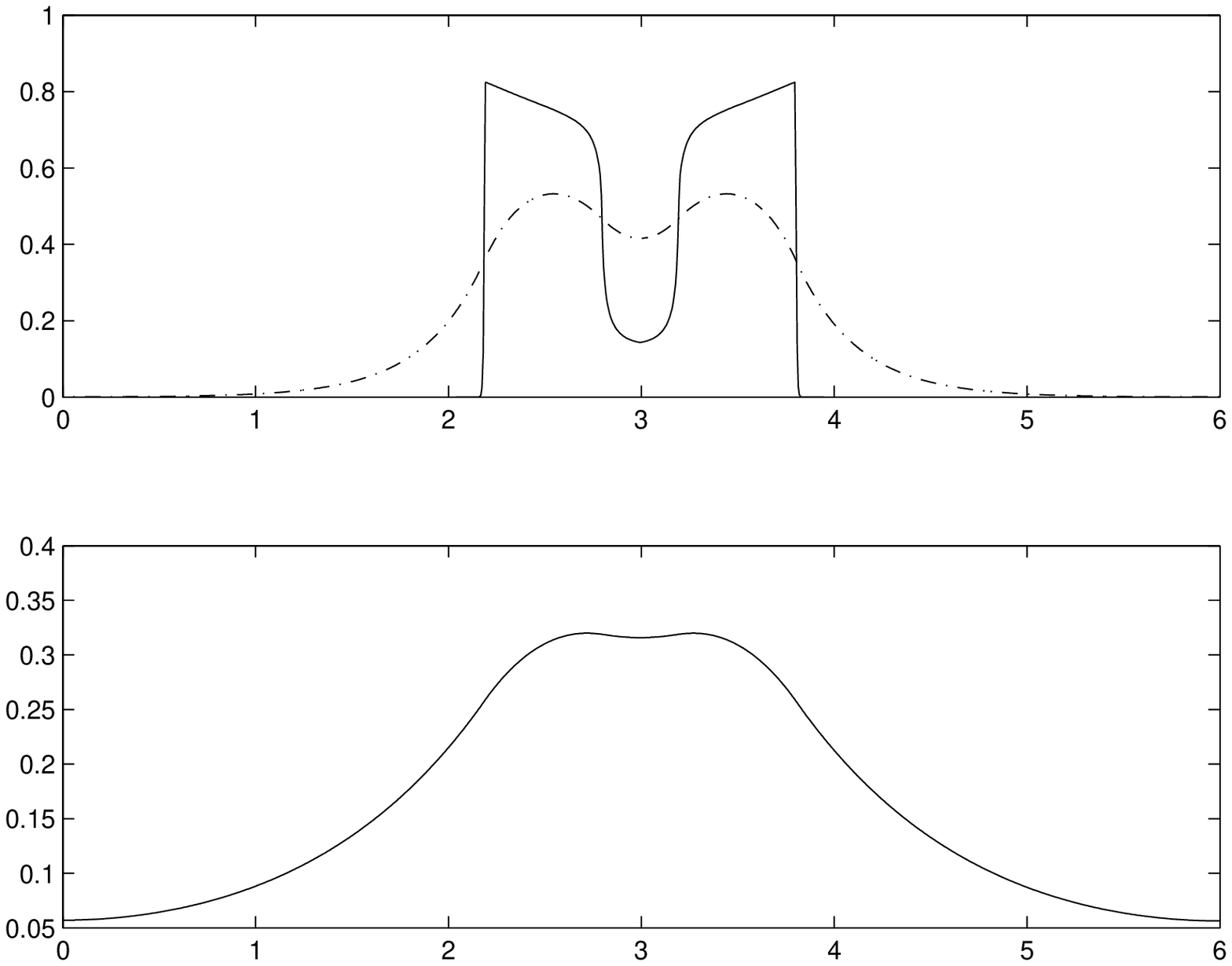}
\includegraphics[width=0.45\textwidth]{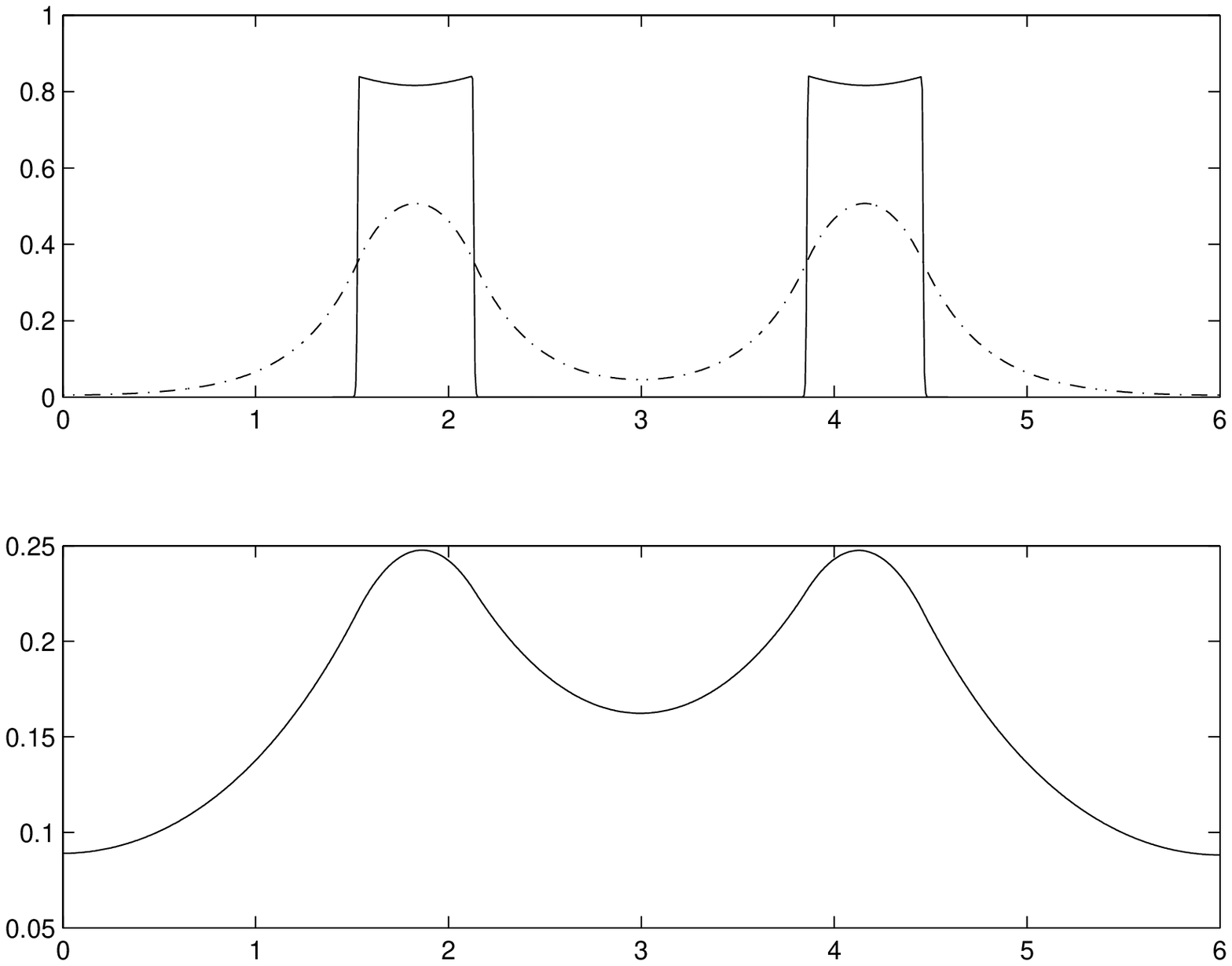}
\vspace{-2mm}
\caption{The instability for $D_c<D_S$ when $M= 1$. We observe that the plateau splits in two pieces, each of which
stabilizes after they are well separated. In these four pairs of figures,
the solid and dashed lines on the top sub-figure represent $n$ and $c$, respectively, and the
bottom sub-figure shows~$S$.}\label{fig:unstable}
\end{center}
\end{figure}

\begin{figure}
\begin{center}
\includegraphics[width=0.45\textwidth]{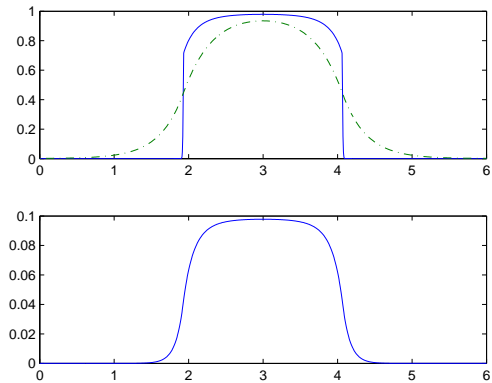}
\vspace{-6mm}
\caption{The stability of the plateau when $D_c>D_S$, even with $M= 2$. The parameters are
$\alpha_c=1,\alpha_S=0.1,D_c=0.1,D_S=0.01$. The solid and dashed lines on the top sub-figure represent $n$ and $c$ respectively and the
bottom sub-figure shows $S$.}\label{fig:steadyLbig}
\end{center}
\end{figure}

We present three families of results in Figures  \ref{fig:steady}, \ref{fig:unstable} and  \ref{fig:steadyLbig}. In these
pictures, the solid and dashed lines on the sub-figures at the top represent $n$ and $c$ respectively, while
the bottom sub-figures depict $S$.

First we illustrate Theorem \ref{th:laststst} (that is $L$ small) in Figure \ref{fig:steady}. Here the computational domain is
$[0,6]$ and the initial density is an almost centered indicator function:
$$
n^0= \I_{[3,3.5]} \,.
$$
We observe the different shapes of the steady states supporting the formal asymptotics above.
When $D_c>D_S$, the positive part of $n$ is concave, when $D_c=D_S$ it is flat, and when $D_c<D_S$ it is convex.
The leading order approximation for small $L$ suggests the relation $M \approx L(1-\gamma)$ between the total mass
$M$ and the width of the plateau. Since $M$ and $L$ are not very small in these computations, this approximative
relation should, however, be corrected by higher order terms. In particular, note that in the simulations leading to the
right picture, $\gamma = 1$ holds, such that the leading order term does not provide any contribution to the total mass.

Secondly, we test for $L$ big. The numerical results for $\alpha_c=1,\alpha_S=1,D_c=0.1,D_S=1$
are presented in Figure \ref{fig:unstable},
with initial data corresponding to $M=1$, namely
$$
n^0(x)=\I_{[2.5,3.5]} \,.
$$
It seems that the total mass is too large for a one-plateau steady state to exist. The initial plateau splits into two
smaller ones which appear to be stable.

It appears numerically that, when $\gamma\leq 1$, $D_c>D_S$, we always reach a steady state solution with concave
cell density in the plateau,
no matter how large the total mass is. As an illustration we show numerical results with $M=2$ and the initial density
$$
n^0(x)=\I_{[2,4]} \,.
$$
The results are shown in Figure \ref{fig:steadyLbig}.

\section{Steady states for almost equal diffusion lengths}
\label{sec:stst2}

Obviously, for $D_c=D_S$, i.e. $L_c=L_S$, and for $\gamma = \alpha_S/\alpha_c < 1$ the steady state problem
(\ref{eq:fp}) has the constant solution $n = n_0 = 1-\gamma$ (for arbitrary $L>0$). Note that, in terms of the original unscaled
parameters (as occurring in (\ref{eq:HYKS})) the equality of the scaled diffusivities means equality of the quantities
$\sqrt{D_c\tau_c}$ and $\sqrt{D_S\tau_S}$, which can be interpreted as diffusion lengths, i.e. the average distance
a molecule diffuses, before it gets degraded.

When considering the dynamics (\ref{eq:red-dyn}) with $D_c=D_S$, it is obvious that $c\alpha_S = S\alpha_c$ holds.
Therefore (\ref{eq:red-dyn}) is equivalent to
\begin{equation} \label{eq:red-dyn1}
\left\{\begin{array}{ll}
\p_tn+\partial_x \left(n(1-\gamma-n)\partial_x c \right)=0 \,,\\[2mm]
-D_c\partial_x^2 c+  c = \alpha_c n \,.
\end{array}
\right.
\end{equation}
For $\gamma<1$ this is exactly the problem analyzed in \cite{DS}, where strong arguments for the stability of one-plateau
solutions are given.

In the remainder of this section we carry out a perturbative analysis to prove existence of a plateau steady state
for small values of $L_S-L_c$, and we illustrate
the qualitative behaviour of the solution by a formal asymptotic expansion in terms of the perturbation parameter.

\subsection*{Existence of a steady state}

Our approach is completely analogous to the preceding section. We fix $L_c>0$, introduce the perturbation parameter
$\delta := L_S-L_c$, and use the decomposition (\ref{eq:decomp}):
$$
\frac{F[L_c+\delta,n]}{1/2-x} = G[n] + \delta g[\delta,n] \,,
$$
with $G[n] = F[L_c,n]/(1/2-x)$ and
$$
g[\delta,n] = \frac{1}{1/2-x} \int_x^{1/2} \varphi(\delta,x,y)n(y)dy + \int_0^x \psi(\delta,x,y)n(y)dy \,,
$$
where
\begin{eqnarray*}
\varphi(\delta,x,y) &=& \frac{e^{(L_c+\delta)(x-y)} + e^{(L_c+\delta)(x+y-1)} - e^{L_c(x-y)} - e^{L_c(x+y-1)}}{\delta} \,,\\
\psi(\delta,x,y) &=& \frac{e^{(L_c+\delta)(x+y-1)} - e^{(L_c+\delta)(y-x)} - e^{L_c(x+y-1)} + e^{L_c(y-x)}}{\delta(1/2-x)} \,.
\end{eqnarray*}
With this notation, the fixed point problem (\ref{eq:fp}) reads
\beq\label{eq:fp2}
 n = 1-\gamma - \delta\gamma \frac{g[\delta,n]}{G[n]} \,.
\eeq
So we need uniform (in $\delta$) boundedness of the linear maps $g$ and $G$, as well as boundedness away from
zero of $G[n]$.

\begin{lemma}
Let  $\gamma = \alpha_s/\alpha_c < 1$ and let
$$
B := \left\{ n \in C([0,1/2]):\ \|n-1+\gamma\|_\infty < C_B \right\} \,,\qquad\mbox{with }
  C_B = \frac{(1-\gamma)e^{-L_c/2}}{2(2 - e^{-L_c/2})} \,.
$$
Then
$$
G[n] \ge \frac{(1-\gamma)e^{-L_c/2}(1-e^{-L_c})}{2L_c} \,,
$$
for $n\in B$ and the map $G$ is bounded with respect to $\|\cdot\|_\infty$.
\end{lemma}

\proof
The boundedness follows immediately from the representation
$$
G[n](x) =  \frac{e^{L_c x}}{1/2-x} \int_x^{1/2} \left(e^{-L_c y}+e^{-L_c(1-y)}\right) n(y)dy
- \frac{e^{-L_cx} -e^{-L_c (1-x)}}{1/2-x} \int_0^x e^{L_c y} n(y)dy \,.
$$
Since $C_B < (1-\gamma)/2$, $n>0$ for $n\in B$. Therefore, for $x\in(0,1/2)$, $G[n]$ is the difference of two positive
terms, and we can estimate
$$
G[n](x) \ge \frac{e^{L_c x}(1-\gamma-C_B)}{1/2-x} \int_x^{1/2} \left(e^{-L_c y}+e^{-L_c(1-y)}\right) dy
- \frac{e^{-L_cx} -e^{-L_c (1-x)}}{1/2-x} (1-\gamma+C_B) \int_0^x e^{L_c y} dy \,.
$$
Evaluation of the integrals leads to
$$
G[n](x) \ge \frac{1}{B\!e(L_c(2x-1))} \left[1-\gamma-C_B - (1-\gamma+C_B)\left(1-e^{-L_cx}\right)\right]  \,,
$$
where the Bernoulli function $B\!e(z) = z/(e^z-1)$ is positive and strictly decreasing. By setting $x=0$ in its argument
and $x=1/2$ in the bracket, the lower bound from the statement of the lemma is achieved.
\qed

\begin{lemma}\label{lm:Lipschitz2}
For $0<\delta \le \bar\delta$, the linear mapping $n\mapsto g[\delta,n]$ on $C([0,1/2])$ is bounded with a bound only
depending on $\bar\delta$ and $L_c$.
\end{lemma}

\proof
The result is a straightforward consequence of the observation that the functions $\varphi$ and $\psi$ are bounded for
$\delta\in (0,\bar\delta]$, $x,y\in [0,1/2)$. This again follows from the facts that their denumerators
$\tilde\varphi(\delta,x,y)$ and, respectively, $\tilde\psi(\delta,x,y)$ are smooth functions of their arguments satisfying
$\tilde\varphi(0,x,y) = \tilde\psi(0,x,y) = \tilde\psi(\delta,1/2,y) = 0$.
\qed\medskip

The last two results immediately imply the contraction property of the right hand side of (\ref{eq:fp2}) acting on $B$ for
small enough $\delta$, which proves the following existence result.

\begin{theorem}\label{th:L1L2}
Assume that $\gamma := \f{\alpha_S}{\alpha_c}<1$ and define $\delta = L(D_S^{-1/2} - D_c^{-1/2})$. For $|\delta|$
small enough (with $L$ and $D_c$ fixed), there exists a unique entropy solution of \eqref{laststst} of the form
$$
n(x)= \left\{ \begin{array}{ll}
0 & \text{for } \; x \notin [0,L],
\\[3mm]
1- \gamma + O(\delta) \quad & \text{for } \; x \in (0,L),
\end{array}\right.
$$
with $n\in C(0,L)$ and symmetric around $x=L/2$.
\end{theorem}

\subsection*{Formal asymptotic expansion -- shape of the plateau}

Similarly to the preceding section we start with the expansion
$$
F[L_c+\delta,n] = F[L_c,n] + \delta \hat F_1[L_c,n] + O(\delta^2) \,,
$$
with
$$
\hat F_1[L_c,n] = \int_0^x e^{L_c(y-x)} (x-y)n(y)dy + \int_x^1 e^{L(x-y)} (x-y)n(y)dy \,.
$$
This leads to the asymptotic expansion for the cell density:
$$
n = 1-\gamma - \delta\gamma \frac{\hat F_1[L_c,1-\gamma]}{F[L_c,1-\gamma]} + O(\delta^2) \,.
$$
A straightforward computation gives
$$
n = 1 - \gamma + \delta\gamma \left( \frac{1}{L_c} +
 \frac{e^{-L_c x}x - e^{-L_c(1-x)}(1-x)}{e^{-L_c x} - e^{-L_c(1-x)}} \right) + O(\delta^2) \,.
$$
Differentiation and the inequality $e^\eta - e^{-\eta} - \eta > 0$ for $\eta>0$ imply that the function in the parentheses
is strictly increasing for $x\in(0,1/2)$ and strictly decreasing for $x\in(1/2,1)$. Therefore $n$ has the same property for
$\delta>0$, i.e. $D_S<D_c$, and the opposite for $D_S>D_c$. These are the same qualitative results as in the preceding
section.

\section{Existence of short traveling plateaus}
\label{sec:tp}

So far, we have proved the existence of stationary solutions, which can be viewed as traveling waves with
zero velocity. In this section, we use a similar route to establish the existence of some non-zero velocity traveling
plateaus. These exist when degradation of the chemo-repellent is neglected, i.e. $\tau_S=\infty$ in \eqref{eq:HYKS}.
With an appropriate nondimensionalization and a reduction to one dimension, the system becomes
\begin{equation}\label{eq:sysL}
\left\{ \begin{array}{l}
 \partial_tn+\partial_x[(1-n)n\partial_xc-n\partial_xS]=0 \,,\\[2mm]
-D_c\partial_{x}^2c+c=\alpha_c n \,,\\[2mm]
\partial_tS-D_S\partial_{x}^2S=\alpha_S n \,.
\end{array} \right.
\end{equation}
We consider plateaus of length $L$ and with speed $\sigma$.
They are defined as functions of the traveling wave variable $x-\sigma t$, which for simplicity is again
denoted by $x$:
\beq\label{eq:sysLtw}
\left\{ \begin{array}{l}
  -\sigma \partial_xn+\partial_x[(1-n)n\partial_xc-n\partial_xS]=0 \,,\\[2mm]
  -D_c\partial_{x}^2c+c=\alpha_c n \,,\\[2mm]
  -\sigma\partial_xS-D_S\partial_{x}^2S=\alpha_S n \,.
\end{array}\right.
\eeq
Again, we restrict our attention to solutions satisfying
$$
n(x)>0 \,,\quad x \in(0,L) \,,\qquad n=0 \,,\quad \mbox{else}.
$$
We rescale space as $x\to Lx$ and the wave speed as $\sigma\to L\sigma$ and obtain for $x\in \R$
\begin{equation}\label{eq:travelL}
\left\{\begin{array}{l}
  -\sigma L^2 \partial_xn+\partial_x[(1-n)n\partial_xc-n\partial_xs]=0 \,,\\[2mm]
  -\frac{D_c}{L^2}\partial_x^2 c + c=\alpha_c n \,,\\[2mm]
  -\sigma \partial_x S - \frac{D_S}{L^2} \partial_x^2 S = \alpha_S n \,,
\end{array}\right.
\end{equation}
with
\begin{equation}\label{eq:lengthL}
n>0 \quad\mbox{in } (0,1) \,,\qquad n=0 \quad\mbox{else}.
\end{equation}
System \eqref{eq:travelL}--\eqref{eq:lengthL} is defined in the whole space $x\in\R$ and we complete it with boundary conditions for $c$ and $S$:
\beq\label{eq:bound}
c(\pm\infty)=0 \,,\qquad S(-\infty)=S_{\infty} \,,\qquad S(+\infty)=0 \,.
\eeq
The wave speed and the far field value $S_\infty$ of the chemo-repellent are considered as unknown and part of the solution.

\begin{theorem}\label{th:TW}
For $\gamma = \frac{\alpha_S D_c}{\alpha_c D_S}< 1$ and $L$ small enough, there is a unique solution of
\eqref{eq:travelL}--\eqref{eq:bound}, such that $n\in W^{1,\infty}(0,1)$ and
$$
n = 1-\gamma + O(L) \quad\mbox{in } (0,1) \,,\qquad \sigma = \frac{\alpha_S (1-\gamma)}{2 D_S} + O(L) \,,\qquad
  S_\infty = 2D_S + O(L) \,.
$$
Furthermore, $c$ is concave in $(0,1)$ and $S$ is non-increasing on $\R$.
\label{th:main}
\end{theorem}

The method of proof extends, with additional technicalities, that of Section \ref{sec:steady}. After integrating the equation
for $n$, we find the following formulas for the solutions of \eqref{eq:travelL}--\eqref{eq:bound}:
\begin{eqnarray}
n&=&1-\frac{\sigma L^2+\partial_xS}{\partial_xc} \,, \qquad x\in [0,1], \label{eq:soluLn} \\
\partial_xc&=& \f{\alpha_cL_c^2}{2}\left( - \int_0^xe^{L_c(y-x)}n(y)dy
  + \int_x^1e^{L_c(x-y)}n(y)dy\right) \,, \qquad x\in \R, \label{eq:soluLc} \\
\partial_xS&=&- \alpha_SL_S^2\int_0^xe^{\sigma L_S^2(y-x)}n(y)dy \,, \qquad x\in \R,\label{eq:soluLs}
\end{eqnarray}
where, again, the notation $L_c = L/\sqrt{D_c}$, $L_S = L/\sqrt{D_S}$ has been used. If the wave speed $\sigma$
was known, the right hand side of \eqref{eq:soluLn} could be (after substitution of \eqref{eq:soluLc}, \eqref{eq:soluLs})
considered as a fixed point operator for the computation of $n$.
The difficulty is that  $\sigma$ is not known a priori. The new ingredient compared to the previous sections is that
$\sigma$ is used to create a zero of the denumerator at the same position where it occurs in the denominator. This
principle to find a traveling wave speed seems to be new.

Formally, the procedure (whose feasibility will have to be proven) is as follows: Given $n$, find $x_0 = x_0[n]$, such
that $F[L_c,n](x_0) = 0$, with $F[L_c,n]$ as defined in (\ref{eq:F}). Then determine $\sigma = \sigma[n]$ such that
\beq \label{eq:sigma}
\sigma = \f{\alpha_S}{D_S}\int_0^{x_0} e^{\sigma L_S^2(y-x_0)}n(y)dy \,.
\eeq
Now the cell density in the plateau can be determined as a fixed point of
\beq\label{eq:FL0}
{\cal F}[n](x) =1- \f{2D_c}{\alpha_c} \,\f{\sigma[n] - \f{\alpha_S}{D_S}\int_0^xe^{\sigma[n] L_S^2(y-x)}n(y)dy}
 {F[L_c,n](x)} \,.
\eeq
The next step is to show that, for $L$ small enough, ${\cal F}[n]$ is well defined on
$$
 B = \left\{ n \in C([0,1]):\ \|n-1+\gamma\|_\infty < \frac{1-\gamma}{2} \right\} \,,
$$
i.e. for $(1-\gamma)/2 =: \underline n < n < \overline n := 3(1-\gamma)/2$.

\begin{lemma} \label{lem:x0}
For $L_c$ small enough and $n\in B$ there exists a unique $x_0 = x_0[n] \in (0,1)$ satisfying $F[L_c,n](x_0) = 0$.
Its dependence on $n$ is Lipschitz:
$$
 |x_0[n_1] - x_0[n_2]| \le C \|n_1 - n_2\|_\infty \,,
$$
for all $n_1,n_2 \in B$, with $C$ independent from $L_c\to 0$.
\end{lemma}

\proof
Existence follows from continuity of $F[L_c,n]$ and from $F[L_c,n](0) > 0$ and $F[L_c,n](1) < 0$. Introducing
$x_j = x_0[n_j]$, $j=1,2$, the equation $F[L_c,n_j](x_j) = 0$ can be written as
$$
\int_0^{x_j} e^{L_c y} n_j(y) dy = e^{2L_c x_j} \int_{x_j}^1 e^{-L_c y} n_j(y) dy \,,\qquad j=1,2\,.
$$
The difference of these two equations is written in the form
\begin{eqnarray*}
&& \int_{x_2}^{x_1} \left( e^{L_c y} + e^{L_c (2x_2-y)} \right) n_1(y)dy
  + \left( e^{2L_c x_2} - e^{2L_c x_1}\right) \int_{x_1}^1 e^{-L_c y} n_1(y)dy \\
&& = \int_0^{x_2} e^{L_c y} (n_2(y)-n_1(y))dy + e^{2L_c x_2} \int_{x_2}^1 e^{-L_c y} (n_1(y)-n_2(y))dy \,.
\end{eqnarray*}
Now the modulus of the left hand side is estimated from below in terms of $|x_1-x_2|$ and the right hand side from
above:
$$
\left( \underline n - 2L_c e^{L_c}\overline n \right) |x_1-x_2| \le 2 e^{L_c} \|n_1-n_2\|_\infty \,.
$$
Since the coefficient on the left hand side can be made positive by choosing $L_c$ small enough, this concludes
the proof.
\qed

\begin{lemma}\label{lem:sigma}
Let the assumptions of Lemma \ref{lem:x0} be satisfied. Then there exists a unique $\sigma = \sigma[n]$ satisfying
\eqref{eq:sigma} with $x_0 = x_0[n]$. Its dependence on $n$ is Lipschitz:
$$
 |\sigma[n_1] - \sigma[n_2]| \le C \|n_1 - n_2\|_\infty \,,
$$
for all $n_1,n_2 \in B$, with $C$ independent from $L_S, L_c\to 0$. Furthermore,
$$
 \frac{\alpha_S (1-\gamma)}{2D_S}e^{-L_S\overline\sigma} =: \underline\sigma \le \sigma \le \overline\sigma
   := \frac{3\alpha_S (1-\gamma)}{2D_S} \,.
$$
\end{lemma}

\proof
Existence and uniqueness of a positive solution follow immediately from the fact that the right hand side of
\eqref{eq:sigma} is positive, bounded, and nonincreasing as a function of $\sigma$. The upper bound is a
consequence of $n\in B$, and so is the lower bound (in the derivation of which the upper bound is also used).
Similarly to the proof of Lemma \ref{lem:x0}, for $n_1,n_2 \in B$ the difference between the corresponding
$\sigma$-equations can be written as
\begin{eqnarray*}
&& \sigma_1 - \sigma_2
  - \frac{\alpha_S}{D_S} \int_0^{x_1} \left( e^{\sigma_1 L_S(y-x_1)} - e^{\sigma_2 L_S(y-x_1)}\right) n_1(y)dy
  = \frac{\alpha_S}{D_S} \int_{x_2}^{x_1} e^{\sigma_2 L_S(y-x_2)} n_1(y)dy \\
&& + \frac{\alpha_S}{D_S} \int_0^{x_1} \left( e^{\sigma_2 L_S(y-x_1)} - e^{\sigma_2 L_S(y-x_2)}\right) n_1(y)dy
    + \frac{\alpha_S}{D_S} \int_0^{x_2} e^{\sigma_2 L_S(y-x_2)} (n_1(y)-n_2(y))dy \,.
\end{eqnarray*}
It is now straightforward to estimate
$$
|\sigma_1 - \sigma_2| \le C\left( |x_1-x_2| + \|n_1-n_2\|_\infty \right) \,,
$$
and to complete the proof by using Lemma \ref{lem:x0}.
\qed

With the definitions of $x_0[n]$ and $\sigma[n]$, the fixed point operator can be written as
$$
{\cal F}[n] = {\cal G}[x_0[n], \sigma[n], n]
$$ 
with notations where the singularity cancelation at $x_0$ appears more clearly
\begin{eqnarray*}
{\cal G}[x_0,\sigma,n] &=& 1-2\gamma \frac{I[x_0,n] + g[x_0,\sigma,n]}{2I[x_0,n] + f[x_0,n]} \,,
\qquad I[x_0,n](x) = \int_x^{x_0} n(y)dy \,,\\
g[x_0,\sigma,n](x) &=& \int_0^{x_0} \left( e^{\sigma L_S^2(y-x_0)} - 1\right)n(y)dy
   - \int_0^x \left( e^{\sigma L_S^2(y-x)} - 1\right)n(y)dy \\
 &=& \int_x^{x_0} \left( e^{\sigma L_S^2(y-x_0)} - 1\right)n(y)dy
   + \int_0^x \left( e^{\sigma L_S^2(y-x_0)} - e^{\sigma L_S^2(y-x)}\right)n(y)dy \,,\\
f[x_0,n](x) &=& F[n](x) - F[n](x_0) - 2 I[x_0,n](x) = \int_x^{x_0} \left( e^{L_c(y-x_0)} + e^{L_c(x_0-y)} - 2\right)n(y)dy \\
&& + \int_0^x \left( e^{L_c(y-x_0)} - e^{L_c(y-x)} \right)n(y)dy
  + \int_x^1 \left( e^{L_c(x-y)} - e^{L_c(x_0-y)} \right)n(y)dy   \,.
\end{eqnarray*}
Since $f$ and $g$ vanish for $L=0$ ($\Rightarrow$ $L_S = L_c = 0$), the constant
$n_0=1-\gamma$ is the only fixed point in this case. The following properties of $I$, $f$ and $g$
are obtained by straightforward computations.

\begin{lemma}
For $n\in B$ there exists a constant $C$ independent from $L\to 0$, such that
\beq\label{eq:I-prop}
 \frac{1}{2}(1-\gamma) \le \frac{I[x_0,n](x)}{x_0-x} \le \frac{3}{2}(1-\gamma) \,,
\eeq\beq\label{eq:g-prop}
\left| g[x_0,\sigma,n](x)-(x-x_0)g[x_0,\sigma,n]'(x)\right| \le CL^2¾(x-x_0)^2 \,,
\eeq\beq\label{eq:f-prop}
\left| f[x_0,n](x)-(x-x_0)f[x_0,n]'(x)\right| \le CL¾(x-x_0)^2 \,,
\eeq\beq\label{eq:der-est}
\left| f[x_0,n]'(x)\right| \le CL \,,\qquad \left|g[x_0,\sigma,n]'(x)\right| \le CL^2 \,.
\eeq
Moreover $C$ only depends on $\|n\|_\infty\leq \bar n$ and not on $n'$.
\end{lemma}
Since we already have the Lemmas
\ref{lem:x0} and \ref{lem:sigma}, we only need to examine the dependence of
${\cal G}[x_0,\sigma,n]$ on its arguments for proving the contraction property of ${\cal F}$. Unfortunately, it will turn out that the Lipschitz constant of
${\cal G}$ as a function of $x_0$ involves the derivative of $n$ with respect to $x$. Therefore,
we shall need a stricter definition of the set, where the fixed point iteration is carried out.

\begin{lemma}\label{lem:selfmap}
There exists a positive constant $\kappa$ such that, for $L$ small enough, the fixed point
operator ${\cal F}$ maps both $B$ into itself and the set
$$
\hat B = \left\{ n\in B:\ n\in W^{1,\infty}([0,1]),\  \|n'\|_\infty \le \kappa L \right\}
$$
into itself.
\end{lemma}

\noindent \proof
We use the alternative representation
\beq\label{eq:fp27}
{\cal F}[n] = 1-\gamma - \gamma \frac{2g[x_0[n],\sigma[n],n] - f[x_0[n],n]}{2I[x_0[n],n] + f[x_0[n],n]}
\eeq
and the controls
\beq\label{eq:fp28}
\frac{I[x_0,n](x) }{x_0-x} \geq \frac{1-\gamma}{2}, \qquad \frac{2I(x)+f(x) }{x_0-x} \geq \frac{1-\gamma}{2}-CL. 
\eeq

As a first step, our previous result, together with the estimate for $\sigma[n]$ in Lemma \ref{lem:sigma}, after
cancellation of $x-x_0$ in the denominator and the denumerator, gives 
$$
 |{\cal F}[n] - 1 + \gamma| \le \frac{CL}{1-\gamma-CL}.
$$
This implies that, for small enough $L$, ${\cal F}$ maps $B$ into itself. The second step is to
compute the $x$-derivative of the fixed point operator:
\beq\label{F-prime}
{\cal G}[x_0,\sigma,n]' = 2\gamma \frac{I'(2g-f) - I(2g'-f') + f'g-g'f}{(2I+f)^2}
\eeq
For estimating this term, we need \eqref{eq:g-prop}, \eqref{eq:f-prop} and the corresponding
property of $I$. This is a first time when the derivative of $n$ enters:
$$
\left| I[x_0,n](x) - (x-x_0)I[x_0,n]'(x)\right| \le \left| \int_x^{x_0} n(y)dy - (x_0-x)n(x)\right| \le
\frac{1}{2}(x-x_0)^2 \| n'\|_\infty \,.
$$
With these properties, for $n\in \hat B$, the modulus of the denumerator in \eqref{F-prime} can
be estimated from above by $CL(1+\kappa L)(x-x_0)^2$. On the other hand, the denominator
can be estimated from below by $(1-\gamma-CL)^2(x-x_0)^2$. Thus, the bound on the derivative
is preserved by ${\cal F}$, if
$$
\frac{C(1+\kappa L)}{(1-\gamma-CL)^2} \le \kappa \,.
$$
This holds for any $\kappa > C(1-\gamma)^{-2}$ and small enough $L$. 
\qed

\begin{lemma}\label{lem:Lipschitz}
For $L$ small enough, $0<x_{01},\,x_{02} < 1$, $\sigma_1$, $\sigma_2$ satisfying the bounds in
Lemma \ref{lem:sigma}, and $n_1,\,n_2\in \hat B$,
$$
 \left\| {\cal G}[x_{01},\sigma_1,n_1] - {\cal G}[x_{02},\sigma_2,n_2] \right\|_\infty \le
 CL \left( \left| x_{01}-x_{02}\right| +  \left| \sigma_1-\sigma_2\right| +
   \left\| n_1-n_2\right\|_\infty\right) \,,
$$
with $C$ independent from $L$. Moreover $C$ only depends on $\|n\|_\infty\leq \bar n$ and not on $n'$.
\end{lemma}

\proof
For analyzing the dependence of ${\cal G}[x_0,\sigma,n]$ on $x_0$, 
it is  convenient to observe the identity ${\cal G}[x_0,\sigma,n](x) = {\cal G}[x,\sigma,n](x_0)$, following from the skew symmetry
of $I[x_0,n](x)$, $f[x_0,n](x)$, and $g[x_0,\sigma,n](x)$ with respect to $x$ and $x_0$. It implies
$$\frac{d}{dx_0}{\cal G}[x_0,\sigma,n](x) = {\cal G}[x,\sigma,n]'(x_0) \,.$$
A bound of the form $CL$ of this quantity has been shown in the proof of the previous lemma.

From the definition of $g[x_0,\sigma,n]$ it is obvious that the derivative with respect to $\sigma$
is $O(L^2)$.

Finally, $I[x_0,\cdot](x)$, $f[x_0,\cdot](x)$ and $g[x_0,\sigma,\cdot](x)$ are linear functionals
with the obvious bounds
$$
\left\| \frac{I[x_0,n]}{x-x_0}\right\|_\infty \le \|n\|_\infty \,,\quad
\left\| \frac{f[x_0,n]}{x-x_0}\right\|_\infty \le CL\|n\|_\infty \,,\quad
\left\| \frac{g[x_0,\sigma,n]}{x-x_0}\right\|_\infty \le CL^2\|n\|_\infty \,.
$$
Considering again  the representation \eqref {eq:fp27} and the lower bound  \eqref {eq:fp28}, the proof is completed.
\qed

\bigskip

Combining the Lemmas \ref{lem:x0}, \ref{lem:sigma}, \ref{lem:selfmap}, \ref{lem:Lipschitz}
shows that, for $L$ small enough, ${\cal F}:\hat B\to\hat B$ is a contraction, completing the
proof of the existence and uniqueness statement of Theorem \ref{th:TW}. The limit as $L\to 0$
of $n$ follows from the form \eqref{eq:fp27} of the fixed point operator. Using this limit in the limit
of the equation $F[L_c,n](x_0)= 0$ shows that the limit of $x_0$ is $1/2$. Then the limit of $\sigma$
is obtained from $L_S\to 0$ in \eqref{eq:sigma}, and the limit of the far-field value $S_\infty$ is derived
by using \eqref{eq:soluLs}:
$$
S_\infty = S(-\infty) = S(0) = -\int_0^\infty \partial_x S(x)dx  \to 2D_S \,.
$$
This completes the proof of Theorem \ref{th:TW}.

\section{Numerical simulation of traveling plateaus}
\label{sec:tp-num}

\subsection*{Algorithm}\label{sec:algorithm}

In the previous section, the existence of traveling wave solutions of system \eqref{eq:sysL} in the form of
short enough cell density plateaus has been proven.
In the following, numerical simulations of system \eqref{eq:sysL} will be presented, indicating the necessity
of the shortness assumption for the dynamic stability of traveling plateaus.

In order to obtain fast enough convergence to a traveling wave, while it moves through the finite computational
domain $[0,A]$, an approximation of the wave, corresponding to the limit $L\to 0$, is used as initial condition.

More precisely, the initial cell density $n(t=0)$ is chosen as
\beq\label{eq:numn}
n_0 = (1-\gamma) \I_{(1,1+L)} \,,
\eeq
where we recall that $\gamma=\frac{\alpha_S D_c}{\alpha_c D_S}$. With this cell density, the maximum
of the chemo-attractant concentration occurs at $x_0=1+L/2$, and the velocity $\sigma$ is obtained by solving 
the nonlinear equation \eqref{eq:sigma}. 
Next, an initial datum $S_0$ for the chemo-repellent concentration is computed by solving the equation for $S$ in \eqref{eq:sysLtw} (see also \eqref{eq:soluLs}).

Since the numerical scheme is restricted to a finite computational domain $[0,A]$, we need to define appropriate boundary conditions. Simulation times $T>0$ are chosen such that the set  $\{ x, \; n(t,x)>0 \}$ stays away from
the boundary points $x=0,A$ for $0\le t \le T$ and, since the equation for $n$ is hyperbolic, it is enough to use zero entering flux  boundary conditions.

For the chemical concentrations, Robin boundary conditions are used, which are satisfied exactly by traveling
plateau solutions:
\begin{equation}\label{eq:condc}
\sqrt{D_c}\, \p_x c(t,0) = c(t,0) \,, \qquad \sqrt{D_c}\, \p_x c(t,A) = -c(t,A) \,,
\end{equation}
\begin{equation}\label{eq:condS}
\p_xS(t,0)=0 \,, \qquad D_S\p_xS(t,A)+\sigma S(t,A) = 0 \,.
\end{equation}
The value of $\sigma$ needs to be updated for each time step. This requires the computation of $\sigma[n]$
as described in the previous section, where the cell density $n$ from the previous time step is used. The 
nonlinear equations for $x_0[n]$ and $\sigma[n]$ are solved by the Newton method.

As in Section \ref{sec:numtest}, the equation for $n$ is discretized by the Enquist-Osher scheme and the
quasi-stationary equation for $c$ by finite differences. For the parabolic equation for $S$ a finite difference space
discretization is used with time implicit treatment of the diffusion term.

\subsection*{Numerical results}

\begin{figure}
\begin{center}
\includegraphics[width=0.3\textwidth]{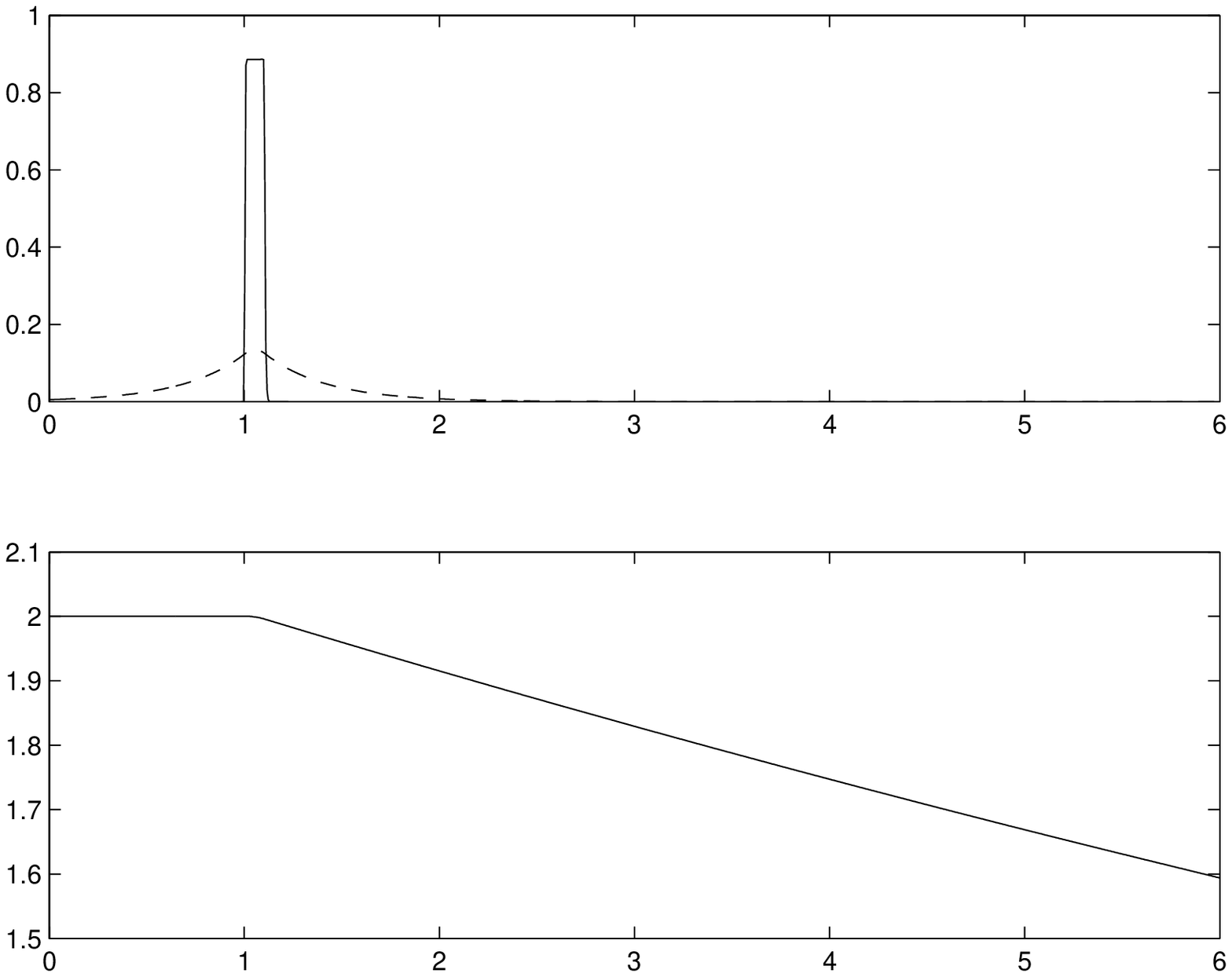}
\includegraphics[width=0.3\textwidth]{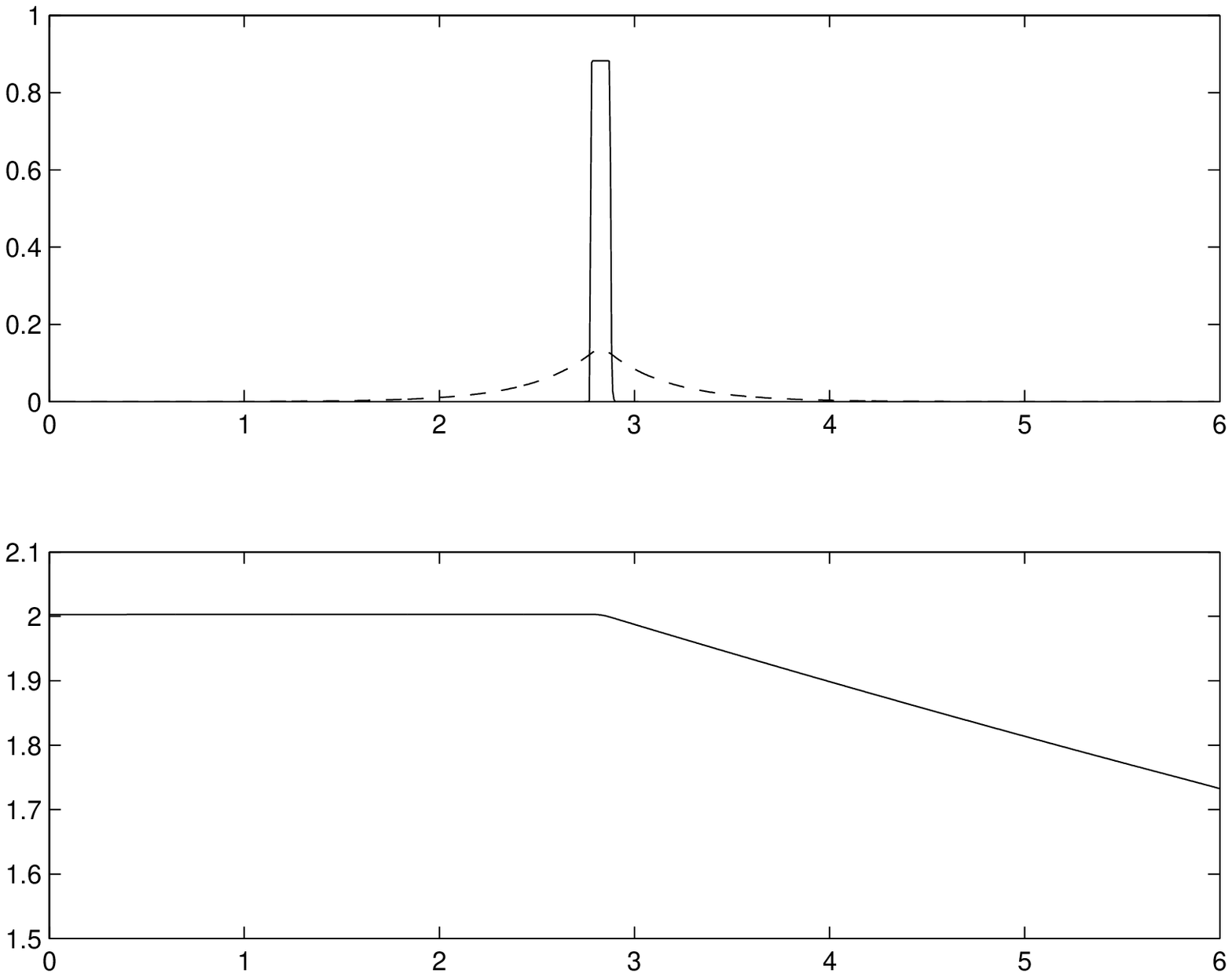}
\includegraphics[width=0.3\textwidth]{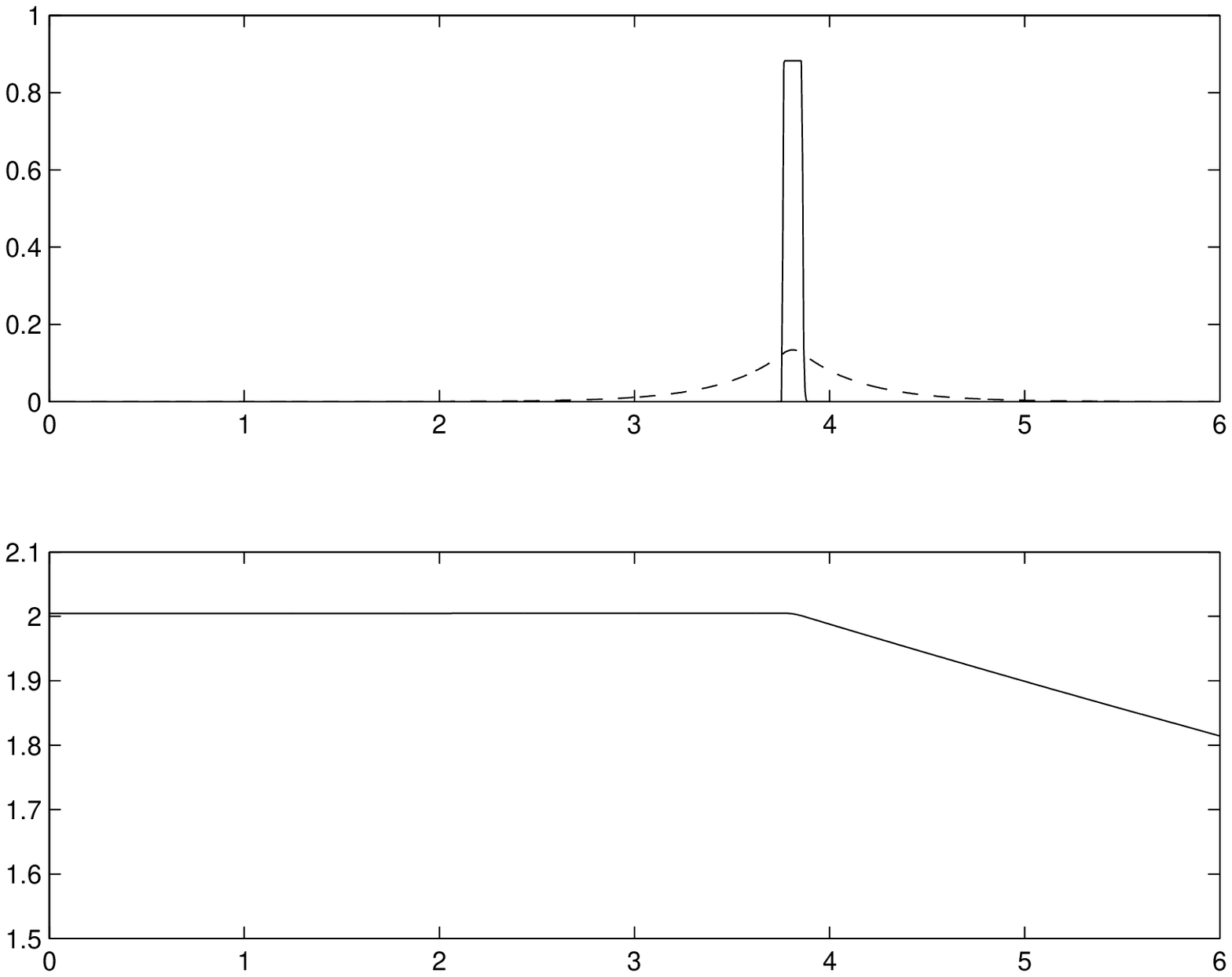}
\vspace{-2mm}
\caption{The solution of (\ref{eq:sysL}) with $L=0.1$ for three different times. Top: The solid and dashed lines are $n$ and $c$, respectively. Bottom:
the evolution of $S$. Left: $t=0$, middle: $t=49.1$, right: $t=81.9$.}\label{fig:TWs01}
\end{center}
\end{figure}

\begin{figure}
\begin{center}
\includegraphics[width=0.3\textwidth]{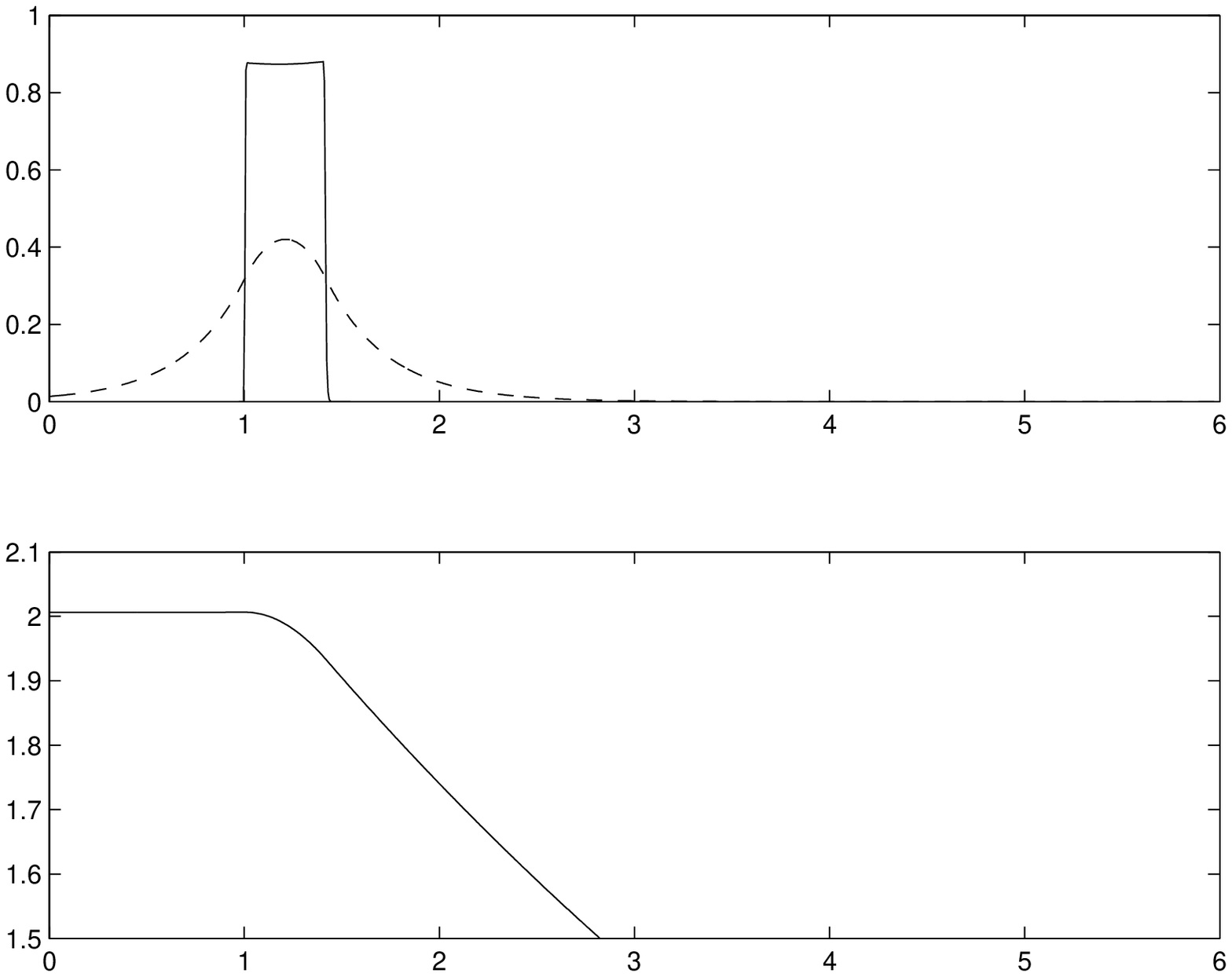}
\includegraphics[width=0.3\textwidth]{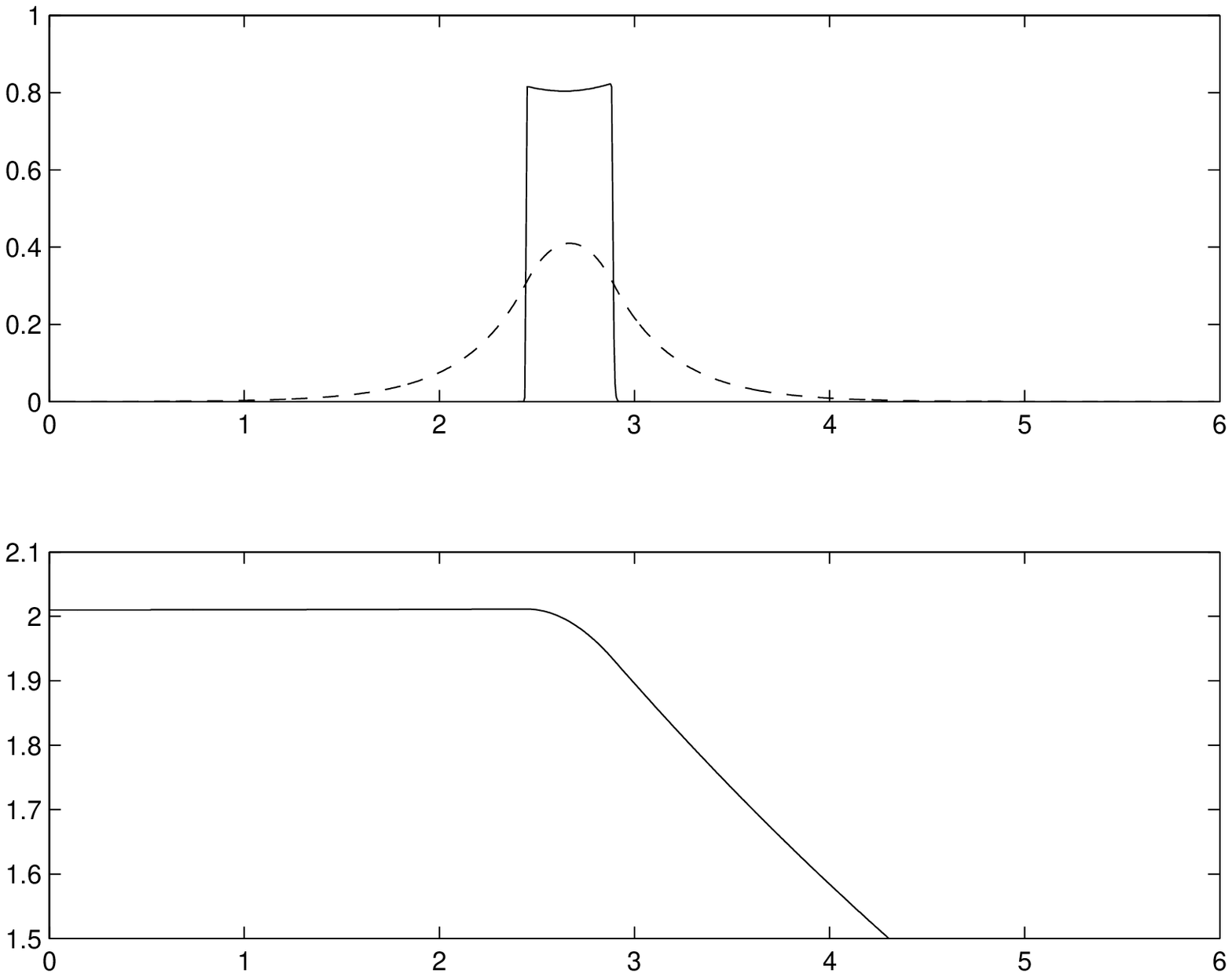}
\includegraphics[width=0.3\textwidth]{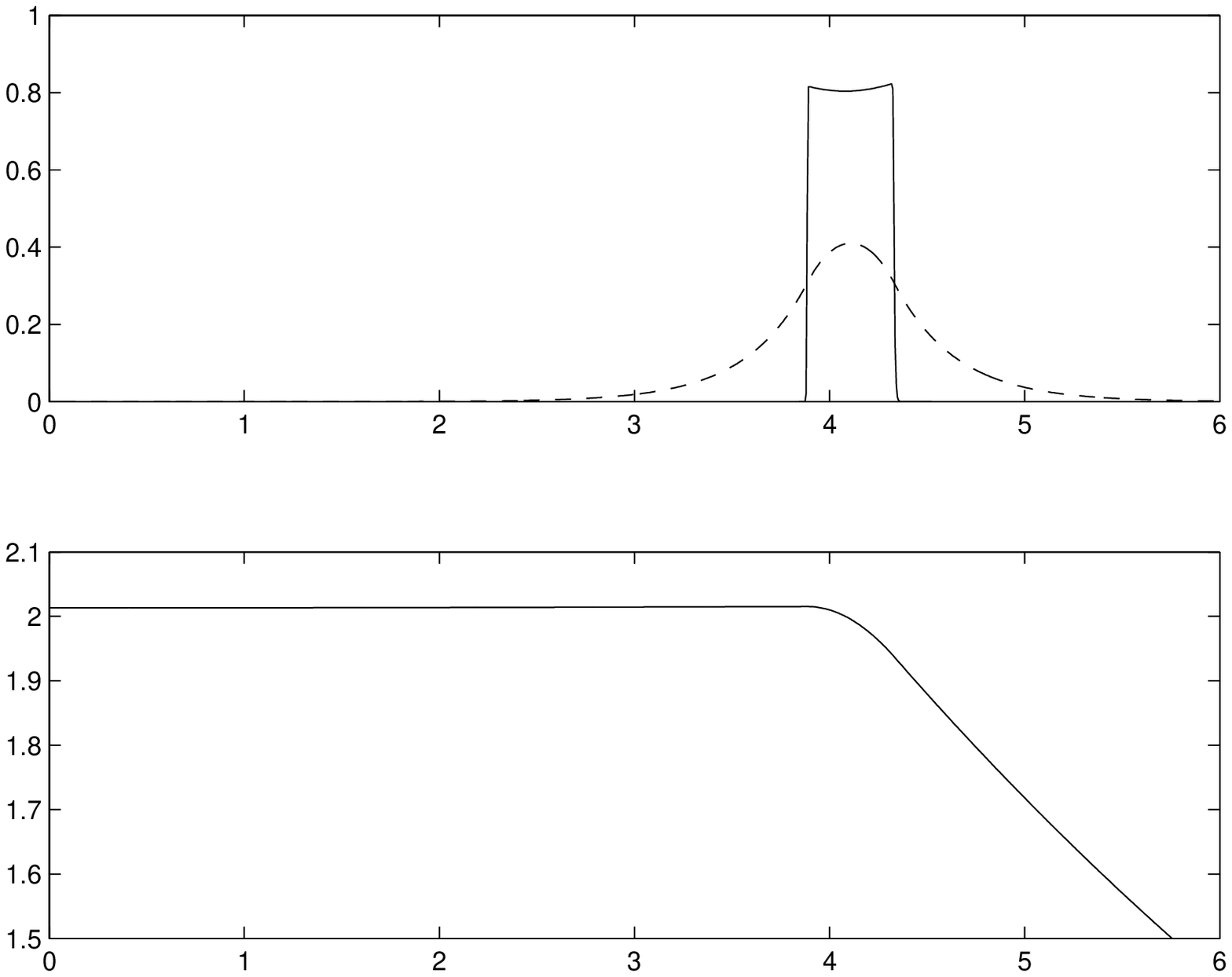}
\vspace{-2mm}
\caption{Same as Figure \ref{fig:TWs01} with  $L=0.4$. The plotting times are now, left: $t=0$, middle: $t=8.1$, right: $t=17.8$.}\label{fig:TWs04}
\end{center}
\end{figure}

\begin{figure}
\begin{center}
\includegraphics[width=0.3\textwidth]{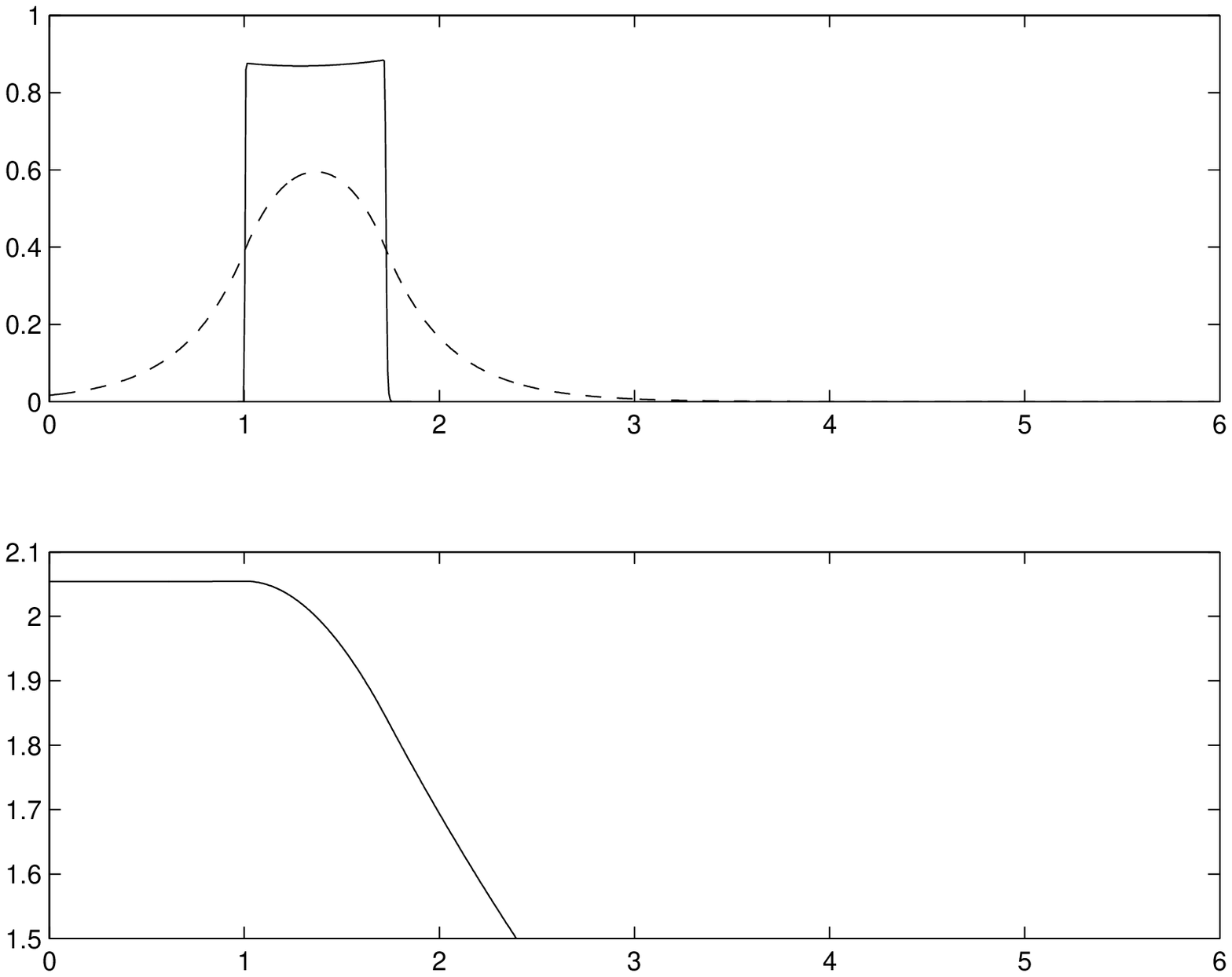}
\includegraphics[width=0.3\textwidth]{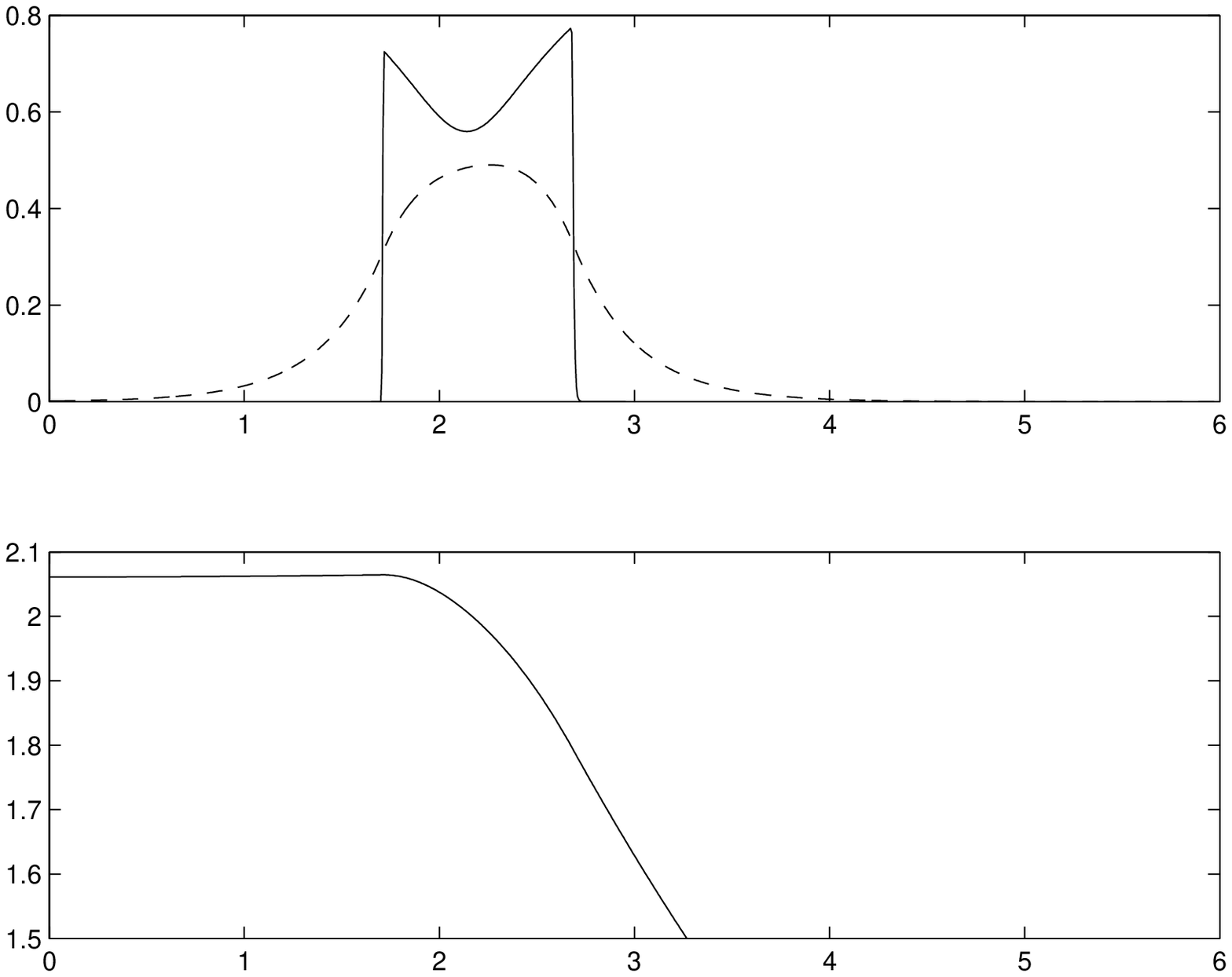}
\includegraphics[width=0.3\textwidth]{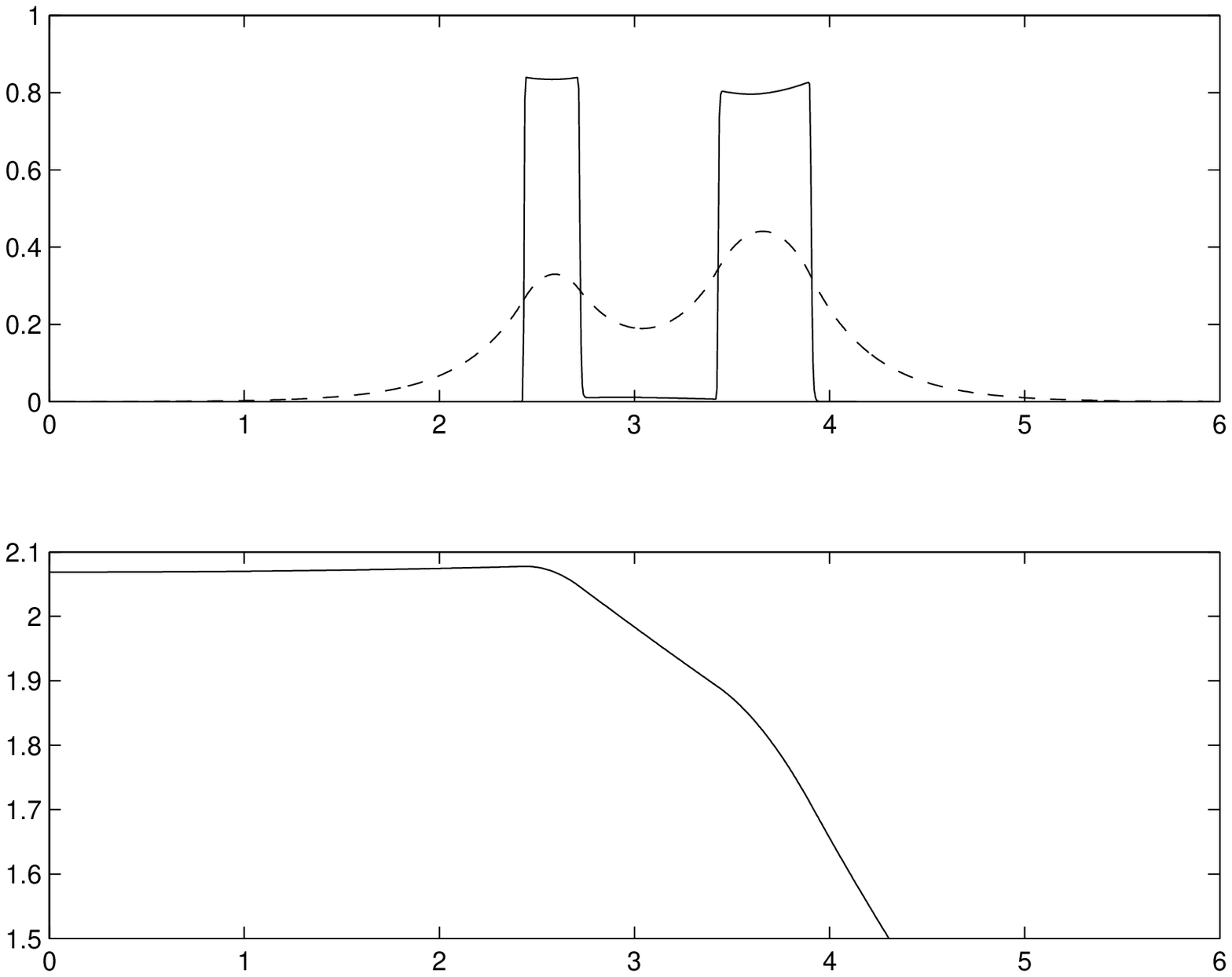}
\vspace{-2mm}
\caption{Same as Figure \ref{fig:TWs01} with $L=0.7$. The plotting times are now, left: $t=0.07$, middle: $t=3.55$, right: $t=6.77$.}\label{fig:TWs07}
\end{center}
\end{figure}

\noindent{\bf Effect of $L$:}  In all our simulations, the computational domain is $[0,6]$, i.e. $A=6$. For the
fixed choice of parameters
\[
\alpha_c=1,\quad  \alpha_S=1, \quad  D_c=0.1, \quad  D_S=1,
\]
we take different values for the length of the initial plateau. For $L=0.1$, $0.4$, and $0.7$, the evolution of
$n$, $c$, and $S$ is plotted in Figures \ref{fig:TWs01}, \ref{fig:TWs04}, and \ref{fig:TWs07}, respectively. We 
observe that, when $L=0.1$ or $0.4$, the solution converges to a plateau traveling wave. Note the differences
in the plotted times between the two figures, indicating the different wave speeds. For the largest value of $L$,  the initial plateau splits into two pieces that travel independently (with speeds dictated by their 
post-splitting lengths). An interesting question (we do not have an answer to) is, whether a dynamically 
unstable traveling wave still exists.
\\

\begin{figure}
\begin{center}
a)\includegraphics[width=0.45\textwidth]{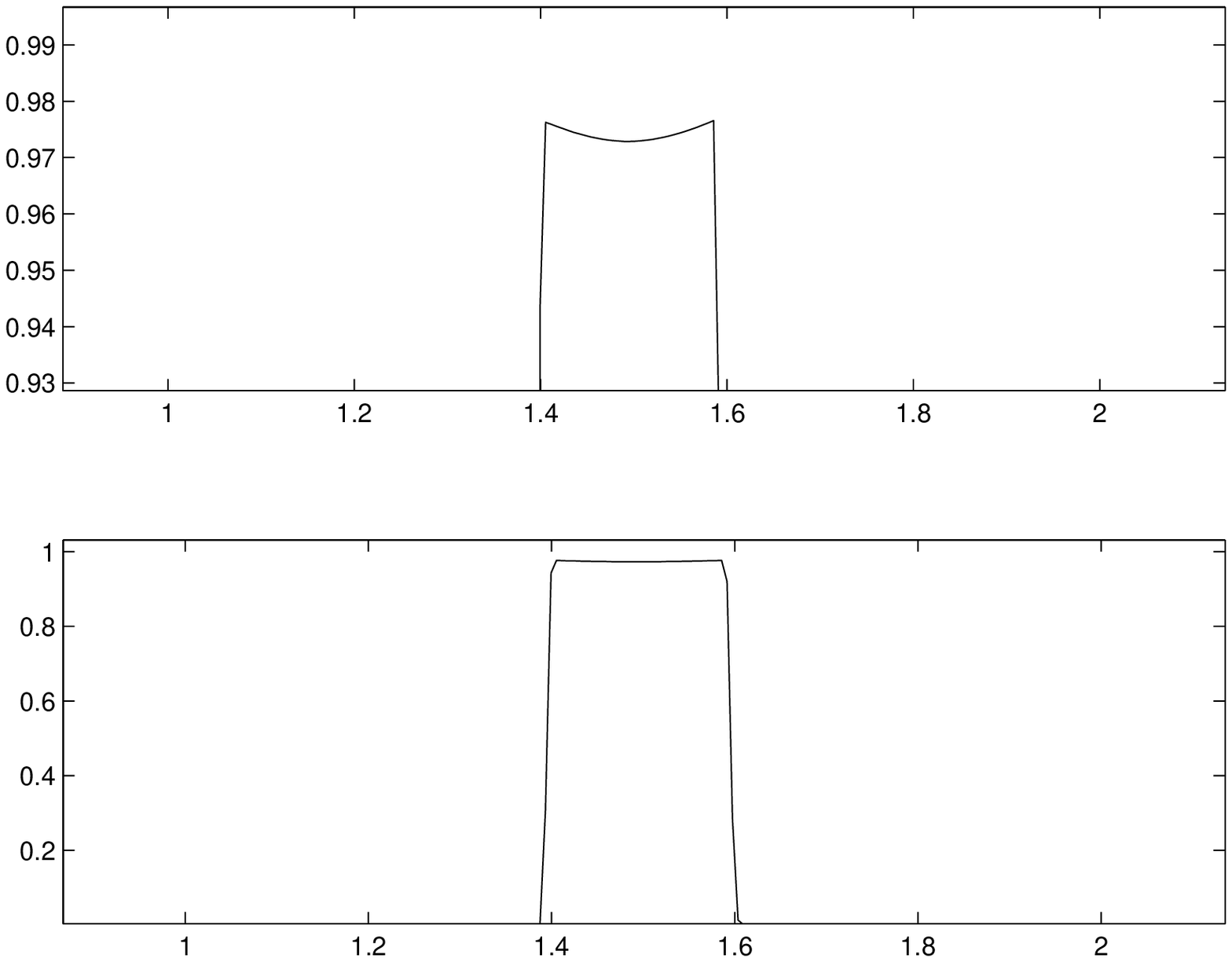}
b)\includegraphics[width=0.45\textwidth]{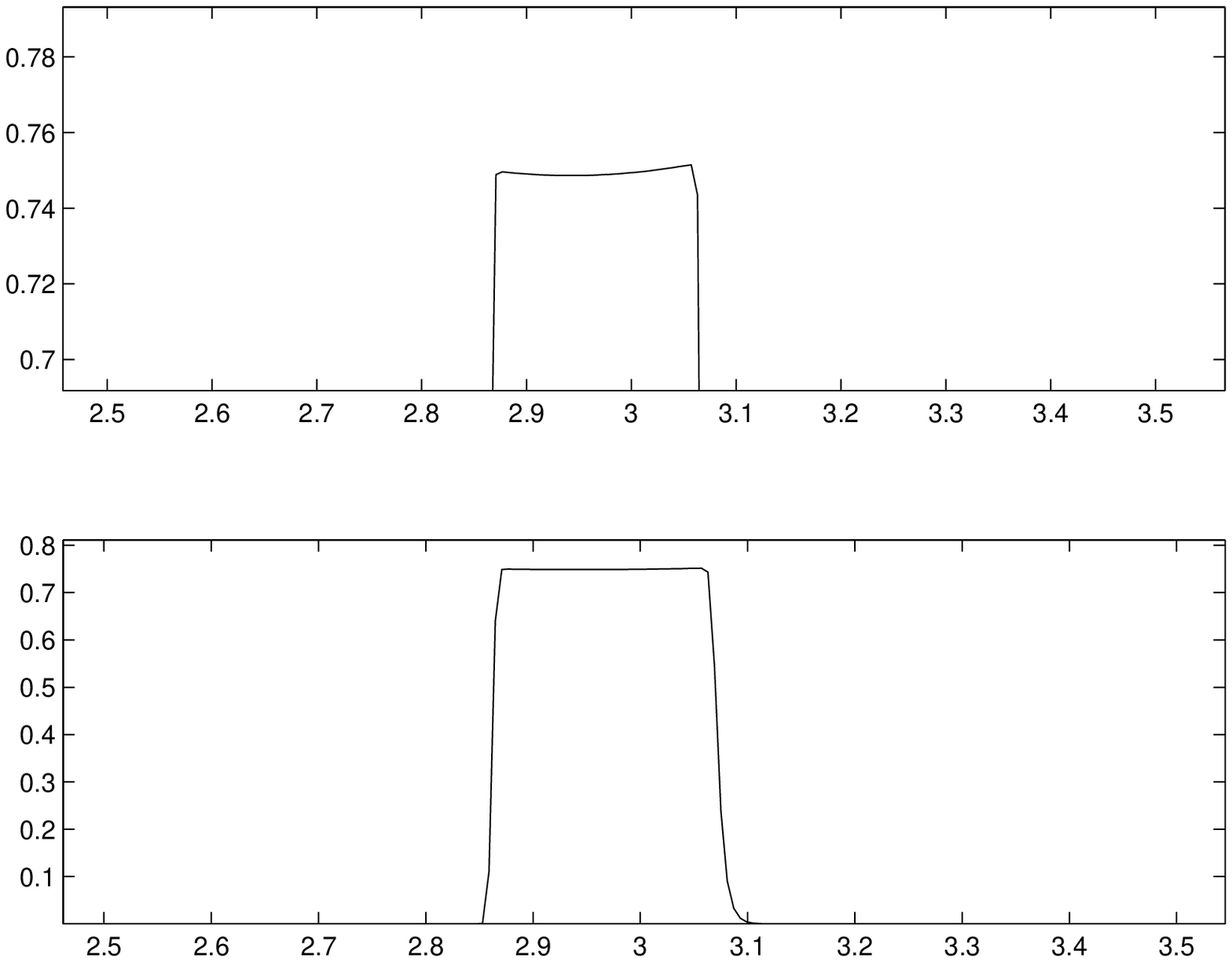}
c)\includegraphics[width=0.45\textwidth]{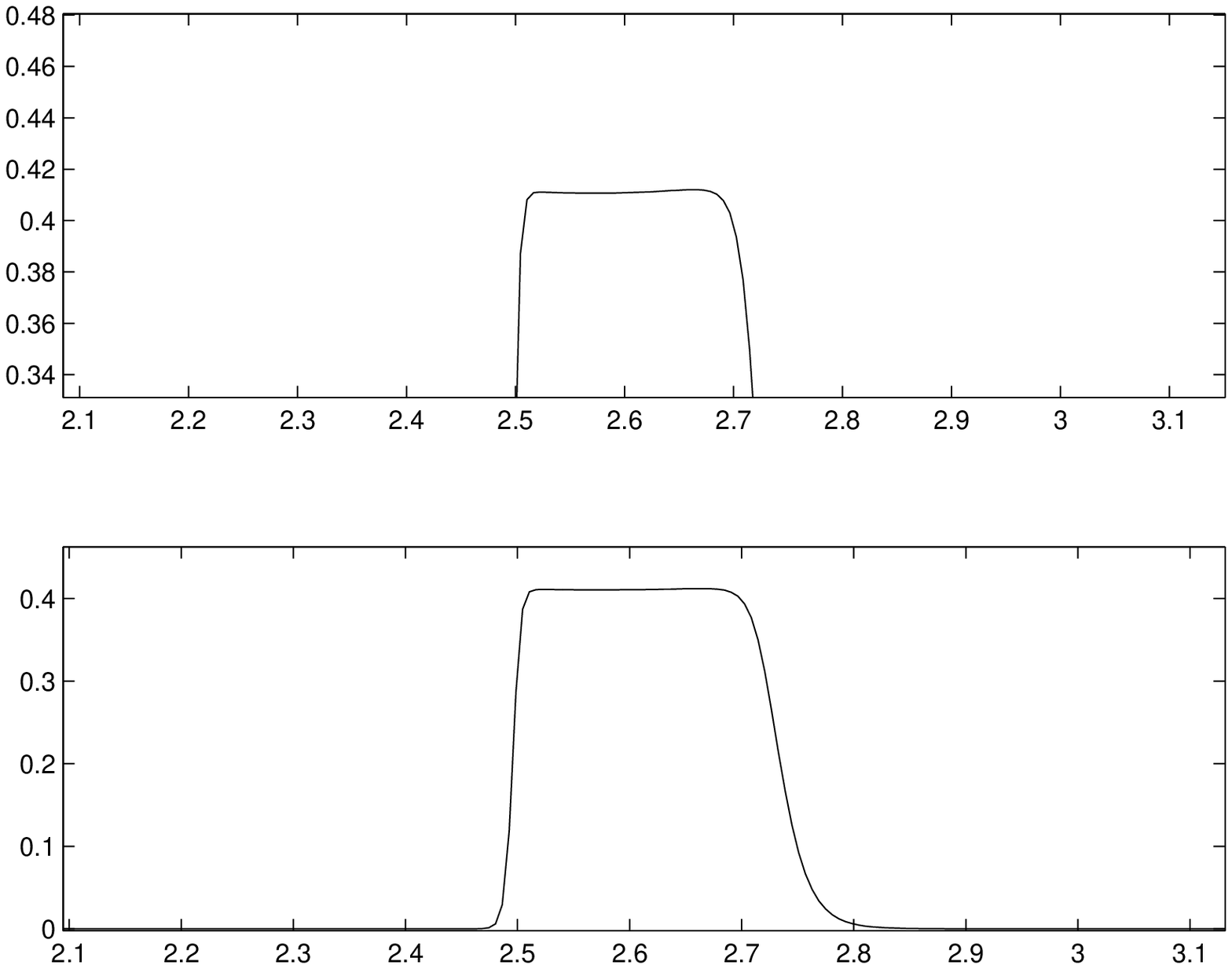}
d)\includegraphics[width=0.45\textwidth]{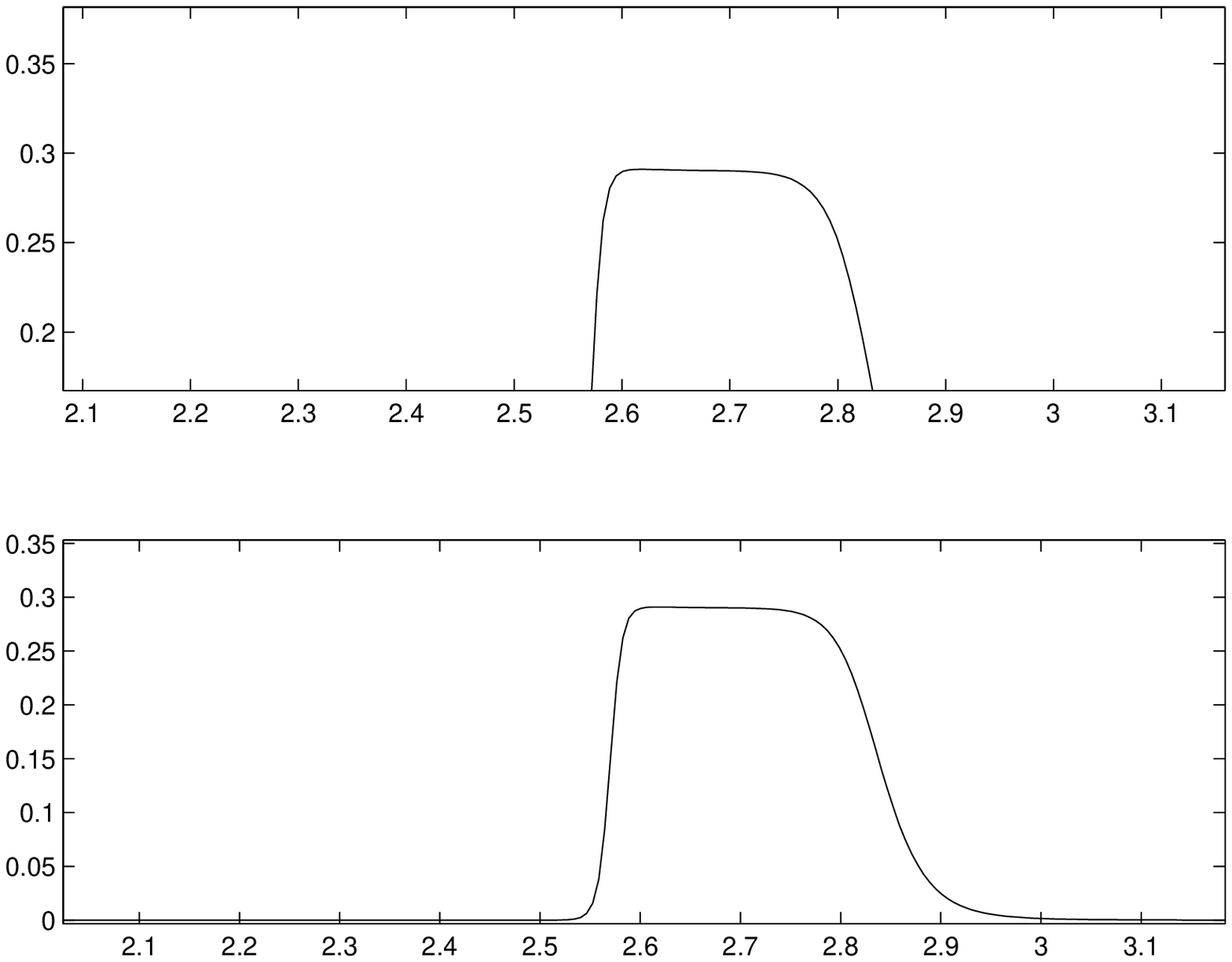}
\vspace{-2mm}
\caption{Different shapes of traveling plateaus. In each figure, the top
subplot depicts a zoom of the peak, while the bottom subplot displays the whole plateau. Parameter values:
$L=0.2$, $\alpha_c=\alpha_S=D_S=1$; a)~$D_c=0.01$, b)~$D_c=0.2$, c)~$D_c=0.5$, d)~$D_c=0.6$.}
\label{fig:TWshape}
\end{center}
\end{figure}

\noindent{\bf Plateau shapes for different physical constants:}
Different parameters can give different shapes for the plateau.
Figure \ref{fig:TWshape} depicts the detailed shape of the plateau 
depending on the diffusion coefficient of the chemical.
It seems that when $D_c$ (and therefore $\gamma$) increases, the plateau 
becomes larger and layers appear on the edges of the plateau.
In this case, attraction forces decrease and therefore cells diffuse more
in the middle. 
\\

\begin{figure}
\begin{center}
\includegraphics[width=0.45\textwidth]{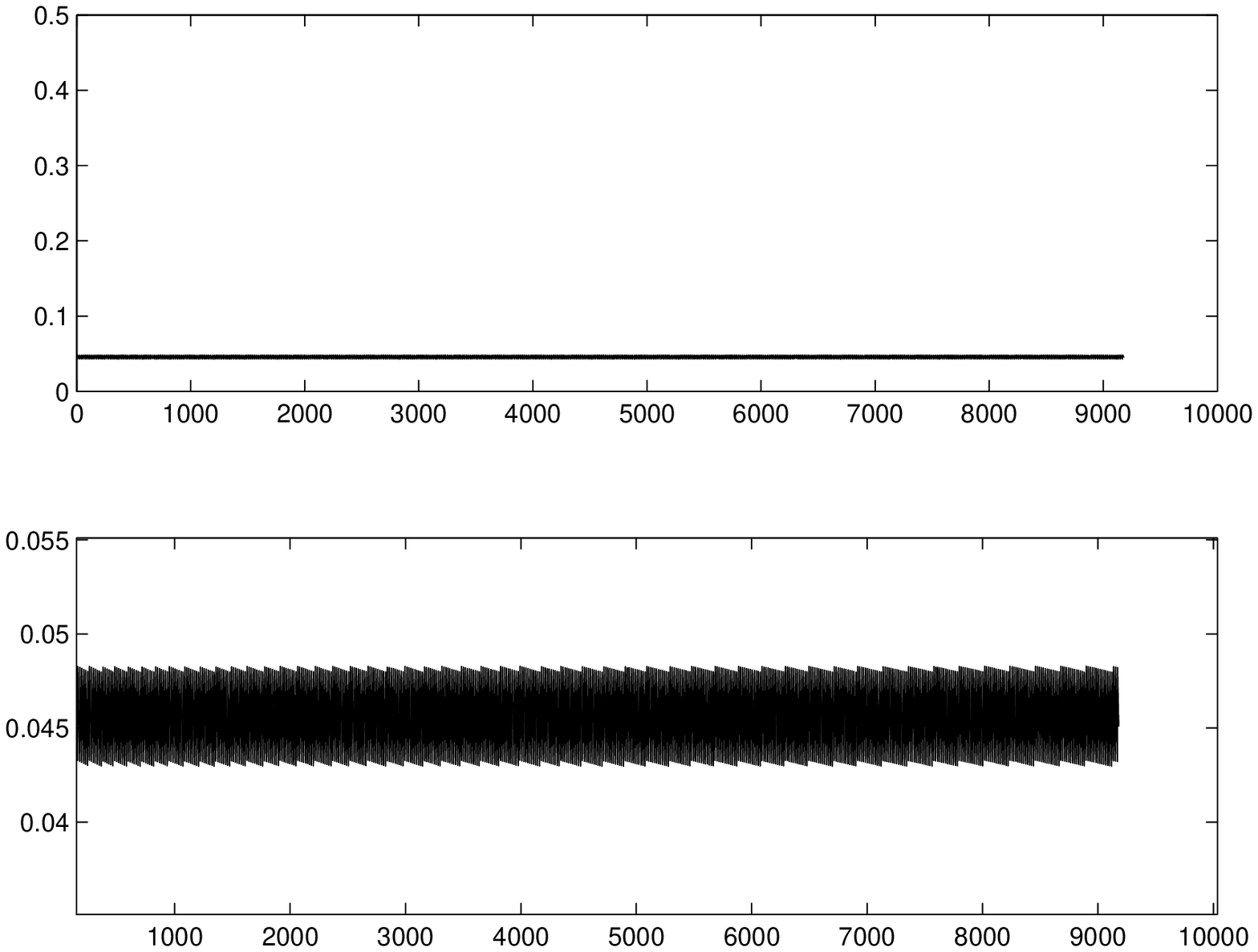}
\includegraphics[width=0.45\textwidth]{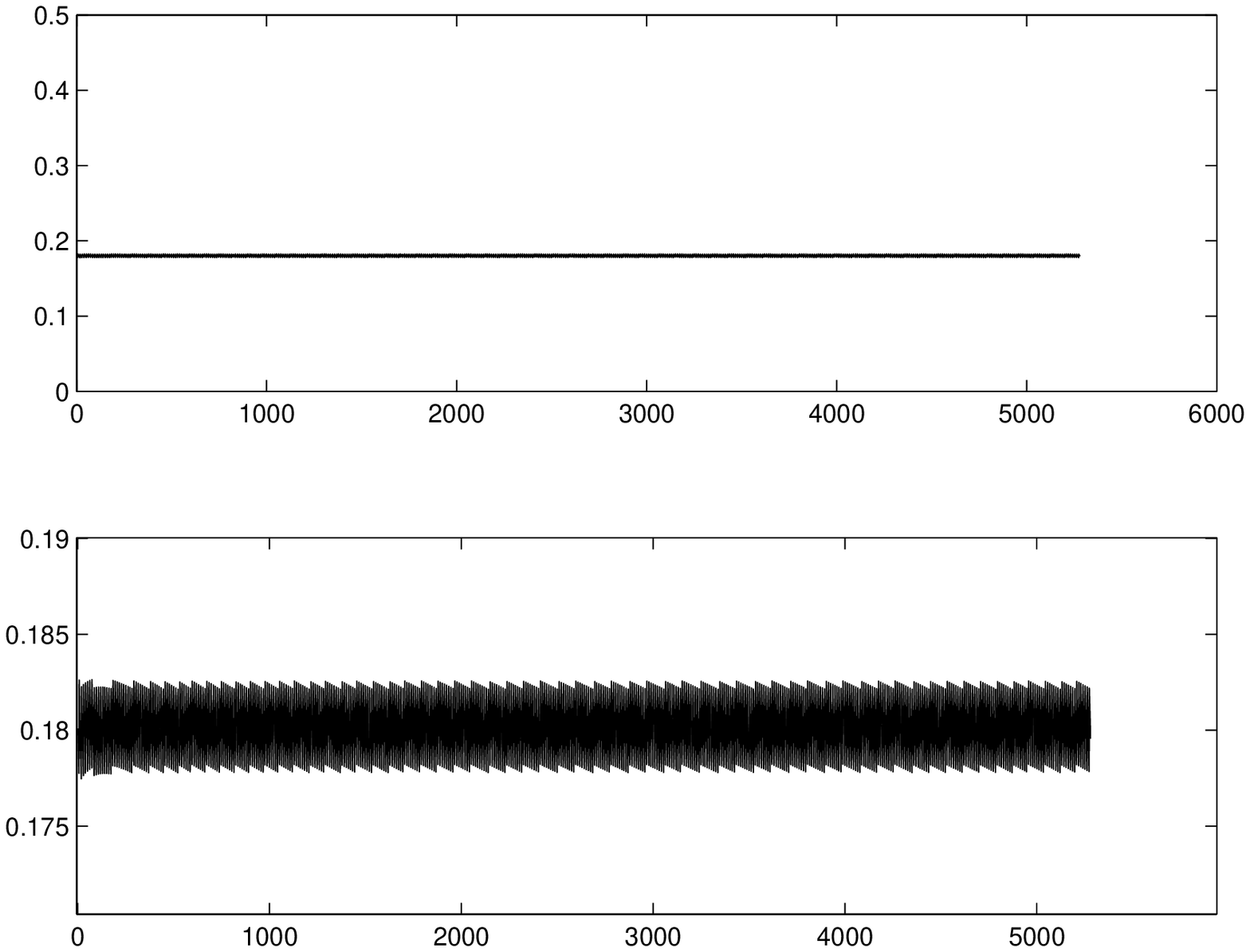}
\vspace{-2mm}
\caption{The velocity of the traveling plateau $\sigma$ for different values of $L$ with the parameters
$\alpha_c=1,\alpha_S=1,D_c=0.1,D_S=1$. We can see that, regardless of the numerical effect,
they are almost constant. Left: $L=0.1$; right: $L=0.4$. Top: the full scale, bottom: zoom on the oscillations.}\label{fig:TWv}
\end{center}
\end{figure}

\noindent{\bf Convergence of the scheme:}
We recall that the wave velocity $\sigma[n]$ is computed in each time step.
The numerical values are displayed in Figure \ref{fig:TWv} for two different values of $L$. It seems that the value
of $\sigma$ computed from the initial data is already very close to its steady state value.
Though there are small oscillations, when the mesh is refined, the amplitude of the oscillations is reduced,
indicating convergence of the numerical method. It is easy to see that, regardless of the oscillations,
$\sigma$ is almost constant. Specifically, its value is close to $\f{\alpha_S}{2D_S}L(1-\gamma)$ in accordance with Theorem \ref{th:TW}.


\bigskip

\noindent{\bf Acknowledgment.} This work was initiated, when C. S. was visiting the INRIA/UPMC team Bang on a visitor position, and completed, when C.S. and M.T. where visitors at the Newton Institute of the
University of Cambridge.


%
%
%

\end{document}